\documentclass[12pt]{article}

\pdfoutput=1

\usepackage[top=80pt,bottom=85pt,left=85pt,right=85pt]{geometry}
\usepackage{amssymb}
\usepackage{amsmath}
\usepackage{graphicx,color,subfigure}
\usepackage{setspace}
\usepackage{sectsty}
\usepackage[debug,pageanchor=false]{hyperref}
\definecolor{link}{rgb}{.8,.15,.1}
\hypersetup{colorlinks=true,linkcolor=link,citecolor=link,urlcolor=link,linktocpage}

\newcommand{\R}{\mathbb{R}}
\newcommand{\C}{\mathbb{C}}
\newcommand{\zz}{\mathbb{Z}} 
\renewcommand{\Re}{\mathrm{Re}}    
\renewcommand{\Im}{\mathrm{Im}}  
\newcommand{\ap}{a}
\newcommand{\am}{a_0}
\renewcommand{\b}{b}
\newcommand{\del}{\partial}
\newcommand{\vol}{\mathrm{vol}}

\makeatletter
\@addtoreset{equation}{section}
\makeatother

\begin{document}

\begin{titlepage}

\begin{center}

\vskip .3in \noindent

{\Large \bf{Supersymmetric AdS$_5$ solutions\\\vspace{.1cm} of massive IIA supergravity}}

\bigskip

Fabio Apruzzi$^1$, Marco Fazzi$^2$, Achilleas Passias$^3$, and Alessandro Tomasiello$^3$\\

\bigskip
{\small 

$^1$ Institut f\"ur Theoretische Physik, Leibniz Universit\"at Hannover, Appelstra\ss e 2, 30167 Hannover, Germany
\\	
\vspace{.1cm}
$^2$ Physique Th\'eorique et Math\'ematique, Universit\'e Libre de Bruxelles, Campus Plaine C.P.~231, B-1050 Bruxelles, Belgium \\ and \\ International Solvay Institutes, Bruxelles, Belgium 
\\
\vspace{.1cm} 
$^3$ Dipartimento di Fisica, Universit\`a di Milano--Bicocca, Piazza della Scienza 3, I-20126 Milano, Italy \\ and \\ INFN, sezione di Milano--Bicocca
	
}

\vskip .3cm
{\small \tt fabio.apruzzi@itp.uni-hannover.de, mfazzi@ulb.ac.be, achilleas.passias@unimib.it, alessandro.tomasiello@unimib.it}
\vskip .6cm
     	{\bf Abstract }
\vskip .1in
\end{center}

Motivated by a recently found class of AdS$_7$ solutions, we classify AdS$_5$ solutions in massive IIA, finding infinitely many new analytical examples. We reduce the general problem to a set of PDEs, determining the local internal metric, which is a fibration over a surface. Under a certain simplifying assumption, we are then able to analytically solve the PDEs and give a complete list of all solutions. Among these, one class is new and regular. These spaces can be related to the AdS$_7$ solutions via a simple universal map for the metric, dilaton and fluxes. The natural interpretation of this map is that the dual CFT$_6$ and CFT$_4$ are related by twisted compactification on a Riemann surface $\Sigma_g$. The ratio of their free energy coefficients is proportional to the Euler characteristic of $\Sigma_g$. As a byproduct, we also find the analytic expression for the AdS$_7$ solutions, which were previously known only numerically. We determine the free energy for simple examples: it is a simple cubic function of the flux integers. 

\noindent

\vfill
\eject

\end{titlepage}

\tableofcontents

\section{Introduction}

The study of supersymmetric conformal field theories (CFT) in four dimensions using holography is by now a venerable subject. Their holographic duals are AdS$_5$ solutions in either IIB supergravity or M-theory.  
A comprehensive analysis of supersymmetric AdS$_5$ solutions of IIB supergravity was carried out in \cite{gauntlett-martelli-sparks-waldram-ads5-IIB}; these include the Freund--Rubin compactifications and the Pilch--Warner solution \cite{Pilch:2000ej}.  Analogous studies were performed for $\mathcal{N} = 1$ \cite{gauntlett-martelli-sparks-waldram-M} and $\mathcal{N} = 2$ \cite{lin-lunin-maldacena} supersymmetric AdS$_5$ backgrounds of M-theory, where new analytic solutions were found. AdS$_5$ solutions arising in M-theory usually have a higher-dimensional origin: they are compactifications (``twisted'' in a certain way) of CFT's in six dimensions. Actually this latter CFT is essentially always the $(2,0)$ theory living on the world-volume of M5-branes, as in  \cite{maldacena-nunez} (and in the more recent examples \cite{gaiotto-maldacena,bah-beem-bobev-wecht}). 

Recently, AdS$_7$ solutions in type II supergravity were classified \cite{afrt}. A new infinite class of solutions was found in massive IIA: the internal space $M_3$ is always topologically an $S^3$, but its shape is not round --- rather, it is a fibration of a round $S^2$ over an interval.\footnote{This Ansatz was also considered in \cite{blaback-danielsson-junghans-vanriet-wrase-zagermann,gautason-junghans-zagermann,blaback-danielsson-junghans-vanriet-wrase-zagermann-2,junghans-schmidt-zagermann}, also in a non-supersymmetric setting.} Both D6's and D8's can be present (and, a bit more exotically, O6's and O8's). The CFT duals of these solutions are $(1,0)$-supersymmetric theories, which were argued in \cite{gaiotto-t-6d} to be the ones obtained in \cite{hanany-zaffaroni-6d,brunner-karch} from NS5-D6-D8 configurations (see also \cite{intriligator-6d,intriligator-6d-II} for earlier related theories). A similar class of $(1,0)$ theories can be found in F-theory \cite{heckman-morrison-vafa,dhtv}.

This prompts the question of whether these $(1,0)$ theories, when compactified on a Riemann surface, can also give rise to CFTs in four dimensions. If so, their duals should be AdS$_5$ solutions in massive IIA. 

In this paper we classify AdS$_5$ solutions of massive IIA, and we find many analytic examples. The new (and physically sensible) ones are in bijective correspondence with the AdS$_7$ solutions; this strongly suggests that their dual CFT$_4$ are indeed twisted compactifications of the $(1,0)$ CFT$_6$. The correspondence is via a simple universal map, which was directly inspired by the map in \cite{rota-t} from AdS$_4$ to AdS$_7$ solutions. At the level of the metric it reads
\begin{equation}\label{eq:intromap}
\begin{split}
e^{2A}(ds^2_{{\rm AdS}_5}+ ds^2_{\Sigma_g})\ + &\ dr^2 + e^{2A} v^2 ds^2_{S^2} \ \to \ \\  
&\sqrt{\frac43}\left(\frac43 e^{2A} ds^2_{{\rm AdS}_7} + dr^2 +\frac{v^2}{1+3v^2}e^{2A}ds^2_{S^2}\right) \ ,
\end{split} 
\end{equation} 
where $A$, $v$ are functions of $r$ and $\Sigma_g$ is a Riemann surface of genus $g\ge 2$. This map is so simple that it also allows us to find analytic expressions for the AdS$_7$ solutions. For example, the simplest massive AdS$_5$ solution has metric
\begin{equation}\label{eq:nice5}
	ds^2 = \sqrt{\frac34}\frac{n_2}{F_0}\left(\sqrt{\tilde y+2}\, (ds^2_{{\rm AdS}_5}+ds^2_{\Sigma_g}) + 
	\frac{d\tilde y^2}{4(1-\tilde y)\sqrt{\tilde y +2}} +\frac19 \frac{(1-\tilde y)(\tilde y+2)^{3/2}}{2-\tilde y} ds^2_{S^2}\right)\ 
\end{equation}
with $\tilde y \in [-2,1]$. Its AdS$_7$ ``mother'', obtained via the map (\ref{eq:intromap}), reads on the other hand
\begin{equation}\label{eq:nice7}
	ds^2= \frac{n_2}{F_0}\left(\frac43\sqrt{\tilde y+2}\,ds^2_{{\rm AdS}_7}+ \frac{d\tilde y^2}{4(1-\tilde y)\sqrt{\tilde y +2}} +\frac13 \frac{(1-\tilde y)(\tilde y+2)^{3/2}}{8 - 4 \tilde y - \tilde y^2} ds^2_{S^2}\right)\ .
\end{equation}
Both these solutions have a stack of $n_2$ D6-branes at $\tilde y=2$, and are regular elsewhere. The D6's can also partially or totally be replaced by several D8-branes, much like in a Myers effect \cite{myers}. (In a way, these solutions realize the vision of \cite{polchinski-strassler}.) Such more complicated solutions are obtained by gluing together copies of (\ref{eq:nice7}), or sometimes also of a more complicated metric that we will see later on. 

We start our analysis in complete generality. We use the time-honored trick of reducing the study of AdS$_5$ solutions to that of Minkowski$_4$ solutions whose internal space $M_6$ has a conical isometry. One can then use the general classification of \cite{gmpt2}, which uses generalized complex geometry on $M_6$. Due to the conical structure of $M_6$, the ``pure spinor equations'' of \cite{gmpt2} become a certain new set of equations on $M_5$. (The idea of applying the pure spinor equations to AdS$_5$ solutions in this way goes back to \cite{gabella-gauntlett-palti-sparks-waldram}, where it was applied to IIB solutions.) It is immediately seen that the only possibility that leads to solutions is that of an SU(2) structure on $M_6$ (where the pure spinors are of so-called type 1 and type 2), which means in turn that there is an identity structure on $M_5$. 

The practical consequence of this is that we can determine the metric on $M_5$ in full generality. It is a fibration of a three-dimensional fiber $M_3$ over a two-dimensional space ${\cal C}$. The three-dimensional fiber also has a Killing vector, which is holographically dual to R-symmetry on the field theory side. The fluxes are also fully determined. The independent functions (one function $a_2$ in the metric, the warping $A$, and the dilaton $\phi$) have to satisfy a total of six PDEs.

The problem simplifies dramatically once we impose what we will call the ``compactification Ansatz''. This consists in imposing that: 1) The metric of $\mathcal{C}$ is conformally related to that of a surface $\Sigma$, which does not depend on the coordinates of the three-dimensional space orthogonal to $\mathcal{C}$ inside $M_5$. The conformal factor is equal to the warping function $e^{2A}$ in front of the AdS$_5$ metric; 2) neither $A$, nor the dilaton $\phi$, nor the function $a_2$ entering the metric and fluxes, depend on the coordinates of $\Sigma$. Under this Ansatz, $\Sigma$ has constant curvature\footnote{For compactifications of $(2,0)$ theories, the fact that $\Sigma$ has constant curvature was explained in \cite{anderson-beem-bobev-rastelli}.} (and we can compactify it to produce a compact Riemann surface $\Sigma_g$); the PDEs reduce to only three. Moreover, these PDEs are all polynomial in one of the local coordinates on $M_3$. Thus they can be in fact reduced to a set of ODEs. At this point the analysis branches out in several possibilities; for each of those, only one ODE survives. In the massless case, there is a ``generic case'', which is the reduction to IIA of the BBBW solution \cite{bah-beem-bobev-wecht0,bah-beem-bobev-wecht}, and two special cases being the reduction of the $\mathcal{N} = 1$ Maldacena--N\'u\~nez solution \cite{maldacena-nunez} and the INST solution \cite{itsios-nunez-sfetsos-thompson}. In the massive case, we get new solutions. Again there is a generic case and two special cases. In the generic case, we find no solutions. The first special case, with positive curvature on $\Sigma_g$,  has singularities which we cannot interpret physically. The second, with constant negative curvature\footnote{Compactifying on $T^2$ the NS5--D6--D8 configurations of \cite{hanany-zaffaroni-6d,brunner-karch} and T-dualizing twice should lead to the NS5--D4--D6 system of \cite{witten-mqcdN2}; the holographic dual to those solutions was found in \cite{aharony-berdichevsky-berkooz}.} on $\Sigma_g$, leads to physically sensible solutions. These latter ones are the main result of this paper. 

Solving the ODE produces several solutions, of which (\ref{eq:nice5}) is the simplest. Without D8's, the most general solution has either two D6 stacks (unlike (\ref{eq:nice5}), which has one), or one D6 stack and one O6. As we already mentioned, there is also the possibility of introducing D8's, which can be done by gluing together copies of (\ref{eq:nice5}), of the Maldacena--N\'u\~nez solution, and possibly also of the more complicated solution we just mentioned. As we also anticipated, the map (\ref{eq:intromap}) can then be used to produce analytical expressions for all the AdS$_7$ solutions in \cite{afrt,gaiotto-t-6d}.

All these new explicit solutions are begging further investigation, particularly regarding their field theory interpretation. This might be the beginning of a correspondence between CFT$_6$ and CFT$_4$ similar to the celebrated class $S$ theories \cite{gaiotto} (although notice that we do not discuss Riemann surfaces with punctures here, as was done in \cite{gaiotto-maldacena}). A feature that those theories also had is that (at the supergravity level) the ratio of the number of degrees of freedom in four and six dimensions is proportional to $g-1$, just like for \cite{gaiotto-maldacena} (and for \cite{bah-beem-bobev-wecht}); this is a simple consequence of the map (\ref{eq:intromap}). We compute the central charges for the CFT$_6$ in a couple of simple cases; for example, for a symmetric solution with two D8's. Along with the NSNS flux integer $N$, there is also another flux integer $\mu$, which is basically the D6 charge of the D8's; the number of degrees of freedom is a simple cubic polynomial in $N$ and $\mu$, and agrees with an earlier approximate computation in \cite{gaiotto-t-6d}. It would be interesting to also compute contributions from stringy corrections, which we have not done here.

This paper is organized as follows. In section \ref{sec:condsusy}, we write the system of pure spinor equations relevant for supersymmetry. In section \ref{generalanalysis}, we analyze the system: we determine the metric and fluxes in terms of a few functions, subject to a set of PDEs, which we summarize in subsection \ref{sub:summary}. In section \ref{compAnsatz} we introduce the compactification Ansatz, for which we are able to give a complete list of cases. One of these classes (apparently the only physically sensible one which was not already known) is then analyzed in \ref{sec:sol} in more detail. The highlights of that analysis are the correspondence to AdS$_7$ in section \ref{sub:ads7}, the explicit solutions in sections \ref{sub:nice}, \ref{sub:gen}, \ref{sub:d8sol}, and the preliminary field theory considerations in section \ref{sub:sum}. In appendix \ref{susyvar} we provide the proof of the existence of a Killing vector on $M_5$. In appendix \ref{simpleAnsatz} we consider an Ansatz simpler than the one in section \ref{compAnsatz}; it reproduces a certain solution of \cite{gauntlett-martelli-sparks-waldram-M}. Finally, in appendix \ref{recoveredsolutions}, we summarize the already known solutions which we recovered in our analysis.

\section{The conditions for supersymmetry}
\label{sec:condsusy}
In this section, we will derive a system of differential equations on forms in five dimensions that is equivalent to preserved supersymmetry for solutions of the type ${\rm AdS}_5 \times M_5$. We will derive it by considering ${\rm AdS}_5$ as a warped product of ${\rm Mink}_4$ and $\R$. We will begin in section \ref{Mink4M6} by reviewing a system equivalent to supersymmetry for ${\rm Mink}_4 \times M_6$. In section \ref{AdS5M5} we will then translate it to a system for ${\rm AdS}_5 \times M_5$.

\subsection{\texorpdfstring{${\rm Mink}_4 \times M_6$}{Mink(4) x M(6)}}
\label{Mink4M6}

Preserved supersymmetry for Mink$_4\times M_6$ was found \cite{gmpt2} to be equivalent to the existence on $M_6$ of an ${\rm SU}(3) \times {\rm SU}(3)$ structure satisfying a set of differential equations. The system is described by a pair of pure spinors 
\begin{equation}\label{eq:phipm}
\phi_- \equiv e^{-A_6}\chi_1^+ \otimes \chi_2^{-\,\dagger} \ ,\qquad 
\phi_+ \equiv e^{-A_6}\chi_1^+ \otimes \chi_2^{+\,\dagger} \ ,
\end{equation}
where the warping function $A_6$ is defined by 
\begin{equation}\label{eq:A6}
ds^2_{10}= e^{2A_6} ds^2_{{\rm Mink}_4} + ds^2_{M_6} \ ,
\end{equation}
and the $\pm$ superscripts indicate the chirality of $\chi_1$ and $\chi_2$. The pure spinors $\phi_-$ and $\phi_+$ can be expressed as a sum of odd and even forms respectively, via application of the Fierz expansion and the Clifford map
\begin{equation}\label{eq:clifford}
dx^{m_1} \wedge \dots \wedge dx^{m_k} \rightarrow \gamma^{m_1 \dots m_k} \ .
\end{equation}

The system of differential equations equivalent to supersymmetry for type IIA supergravity reads:
\begin{subequations}\label{eq:46}
\begin{align}
d_H \bigl(e^{2A_6-\phi}\Re\phi_-\bigr) &= - \frac{c_-}{16} F \ ,
\label{eq:46R1} \\
d_H \bigl(e^{3A_6-\phi}\phi_+\bigr) &= 0 \ , 
\label{eq:46I1} \\
d_H \bigl(e^{4A_6-\phi}\Im\phi_-\bigr) &= - \frac{c_+ e^{4A_6}}{16} *_6 \lambda F \ .
\label{eq:462}
\end{align}
\end{subequations}
Here, $\phi$ is the dilaton, $d_H=d-H\wedge$ is the twisted exterior derivative and $c_{\pm}$ are constants such that 
\begin{equation}
\| \chi_1 \|^2 \pm \| \chi_2 \|^2 = c_{\pm} e^{\pm A_6} \ .
\end{equation}
$F$ is the internal Ramond-Ramond flux which determines the external flux via self-duality:
\begin{equation}\label{eq:F*F}
F_{(10)} \equiv F + e^{6A_6} \vol_4 \wedge *_6 \lambda F \ .
\end{equation}
$\lambda$ is an operator acting on a $p$-form $F_p$ as $\lambda F_p = (-1)^{\left[\frac{p}{2}\right]} F_p$, where square brackets denote the integer part.

\subsection{\texorpdfstring{${\rm AdS}_5\times M_5$}{AdS(5) x M(5)}} 
\label{AdS5M5}

As we anticipated, we will now use the fact that anti-de Sitter space can be treated as a warped product of Minkowski space with a line. We would like to classify solutions of the type AdS$_5 \times M_5$. These in general will have a metric\footnote{Here $ds^2_{{\rm AdS}_5}$ is the unit radius metric on AdS$_5$.} 
\begin{equation}\label{eq:met55}
ds^2_{10} = e^{2 A} ds^2_{{\rm AdS}_5} + ds^2_{M_5} \ .
\end{equation}
Since
\begin{equation}\label{eq:ads5}
ds^2_{{\rm AdS}_5} = \frac{d \rho^2}{\rho^2}  + \rho^2 ds^2_{{\rm Mink}_4} \ ,
\end{equation}
$ds^2_{10}$ in equation \eqref{eq:met55} can be put in the form of equation \eqref{eq:A6} if we take
\begin{equation}\label{eq:43}
e^{A_6} = \rho e^A \ , \qquad ds^2_{M_6} = \frac{e^{2A}}{\rho^2}d \rho^2 + ds^2_{M_5} \ .
\end{equation}

In order to preserve the SO$(4,2)$ invariance of AdS$_5$, $A$ should be a function of $M_5$. In addition, the fluxes $F$ and $H$, which in subsection \ref{Mink4M6} were arbitrary forms on $M_6$, should now be forms on $M_5$. For IIA, $F = F_0 + F_2 + F_4 + F_6$; in order not to break SO$(4,2)$, we impose $F_6=0$. 

Following the decomposition of the geometry of $M_6$ we wish to decompose the system of equations \eqref{eq:46} so as to obtain the system equivalent to preserved supersymmetry for AdS$_5 \times M_5$. We start by decomposing the generators of Cliff$(6)$ as
\begin{equation}\label{6dimgambasis}
	\gamma^{(6)}_\rho  = \frac{e^A}{\rho}1 \otimes \sigma_1 \ , \qquad 
	\gamma^{(6)}_{m}= \gamma_m \otimes \sigma_2 \ , \qquad m = 1,\dots,5
\end{equation}
where $\sigma_1$, $\sigma_2$ are the Pauli matrices and $\gamma_m$ generate  Cliff$(5)$. Accordingly, the chirality matrix $\gamma^{(6)}_7 = 1 \otimes \sigma_3$ and the chiral spinors $\chi^+_1$, $\chi^-_2$ are decomposed in terms of Spin$(5)$ spinors $\eta_1$, $\eta_2$ as
\begin{equation}\label{spindec}
	\chi_1^+ = \sqrt{\frac{\rho}{2}} \, \eta_1 \otimes \begin{pmatrix} 1 \\ 0 \end{pmatrix} \ , \qquad
	\chi_2^- = \sqrt{\frac{\rho}{2}} \, \eta_2 \otimes \begin{pmatrix} 0 \\ 1 \end{pmatrix} \ .
\end{equation}
$\phi_-$ and $\phi_+$ now read
\begin{equation}\label{purespindec}
\phi_- = \frac{1}{2} \left( \frac{e^{A}}{\rho} d \rho \wedge \psi^1_+ + i\psi^1_- \right) \ , \qquad
\phi_+ = \frac{1}{2} \left(-i\frac{e^{A}}{\rho} d \rho \wedge \psi^2_- + \psi^2_+ \right) \ ,
\end{equation}
where
\begin{equation}\label{eq:psi12}
\psi^1 \equiv e^{-A} \eta_1 \otimes \eta_2^\dagger \ , \qquad
\psi^2 \equiv e^{-A} \eta_1 \otimes \overline{\eta_2} \ .
\end{equation}
The bar is defined as $\overline{\eta} \equiv (\eta^c)^\dagger \equiv (B \eta^*)^\dagger = - \eta^t B$, where $B$ is a conjugation matrix that in five Euclidean dimensions can be taken to satisfy $B^*=B$, $B^t=-B$, $B^2=BB^*=-1$. The subscripts plus and minus on $\psi^1$, $\psi^2$ refer to taking the even and odd form part respectively, in their expansion as forms.  One should keep in mind here a comment about odd dimensions: the Clifford map (\ref{eq:clifford}) is not injective. Rather, a form $\omega$ and its cousin $* \lambda \omega$ are mapped to the same bispinor (recall the definition of $\lambda$ right after (\ref{eq:F*F})). Thus a bispinor can always be expressed both as an even and as an odd form, and in particular we have
\begin{equation}\label{eq:*l}
	\psi^{1,2}_- = *\lambda \psi^{1,2}_+ \ .
\end{equation} 

Applying the decomposition \eqref{purespindec} to equations \eqref{eq:46} we obtain a \emph{necessary and sufficient} system of equations for supersymmetric AdS$_5 \times M_5$ solutions:
\begin{subequations}\label{susyeq}
\begin{align}
d_H \bigl(e^{3A-\phi}\Re\psi^1_{+}\bigr) + 2 e^{2A-\phi} \Im\psi^1_- \ &= \ 0 \ , 
\label{susyeqa} \\[.2cm] 
d_H \bigl(e^{4A-\phi}\psi^2_{-}\bigr) - 3i e^{3A-\phi} \psi^2_+ \ &= \ 0 \ ,
\label{susyeqb} \\[.2cm]
d_H \bigl(e^{4A-\phi} \Re\psi^1_-\bigr)\ &= \ 0 \ , 
\label{susyeqc} \\[.2cm]
d_H \bigl(e^{5A-\phi} \Im\psi^1_{+}\bigr) -4 e^{4A-\phi} \Re \psi^1_- \ &= \ \frac{c_+}{8} e^{5A} * \lambda F \ . 
\label{susyeqd}
\end{align}
\end{subequations}
We also obtain the condition $c_- = 0$; it follows that the relation $\| \chi_1 \|^2 \pm \| \chi_2 \|^2 = c_{\pm} e^{\pm A_6}$ becomes
\begin{equation}\label{eq:eta12}
\| \eta_1 \|^2 = \| \eta_2 \|^2 = \frac{1}{2} c_+ e^A \ .
\end{equation}
Henceforth, without loss of generality, we set $c_+ =  2$.

The stabilizer group $\mathcal{G} \in \mathrm{Spin}(5)$ of $\eta_1$ and $\eta_2$ can be either SU(2) or the identity group. In the next section we parametrize $\psi^1$, $\psi^2$ in terms of these structures. We will see however that only the identity case leads to supersymmetric solutions. An identity structure is actually a choice of vielbein; so we will end up parameterizing the $\psi^1$ and $\psi^2$ in terms of a vielbein.

\subsection{\texorpdfstring{Parametrization of $\psi^1$, $\psi^2$ and the identity structure}{Parametrization of psi**1, psi**2 and the identity structure}}
\label{parametrization}

We first consider the case where there is only one spinor, $\eta_1=\eta_2$ of norm $e^{\frac{A}{2}}$. In five dimensions it defines an SU(2) structure. This can be read off from the Fierz expansions of $\eta_1 \otimes \eta_1^\dagger$ and $\eta_1 \otimes \overline{\eta_1}$, which as remarked in (\ref{eq:*l}) can be written both as even and as odd forms:     
\begin{equation}\label{eq:psisu2}
\begin{split}
	&\psi^1_+= \frac14 e^{-ij} \ ,\qquad \psi^2_+= \frac14 \omega \ , \\
	&\psi^1_- = \frac14 v \wedge e^{-ij} \ ,\qquad \psi^2_- = \frac14 v \wedge \omega \ .	
\end{split}
\end{equation}
Application of Fierz identities yields 
\begin{equation}\label{eq:veta}
	v \eta_1 = \eta_1
\end{equation}
and the following set of algebraic constraints on the 1-form $v$ and 2-forms $j$ and $\omega$: 
\begin{align}\label{eq:SU2}
\iota_v v = 1 \ &, \qquad \iota_v j = \iota_v \omega = 0 \nonumber \\
j \wedge \omega = 0 \ &, \qquad
\omega \wedge \omega = 0 \ , \qquad
\omega \wedge \overline{\omega} = 2 j \wedge j = \vol_4,
\end{align}
where $\vol_4$ is the volume form on the four-dimensional subspace orthogonal to $v$. This set of forms and constraints define precisely an SU$(2)$ structure in five dimensions.

In this case, however, the two-form part of (\ref{susyeqb}) tells us $\psi^2=0$, which is only possible for $\eta_1 = 0$. Hence, there are no supersymmetric $\mathrm{AdS}_5 \times M_5$ solutions in type IIA supergravity with an $SU(2)$ structure on $M_5$.

Let us then consider the case of two spinors $\eta_1$ and $\eta_2$, which as mentioned earlier define an identity structure. We can expand $\eta_2$ in terms of $\eta_1$ as 
\begin{equation}\label{eq:eta2}
\eta_2 = \ap \eta_1 + \am \eta_1^c + \frac{1}{2} \b \overline{w} \, \eta_1 \ ,
\end{equation}
where $\ap, \, \am \in \C$, $\b \in \R$ and $w$ is a complex vector that we normalize such that $w\cdot \overline{w} =2$ (so that ${\rm Re} w$ and ${\rm Im} w$ are orthogonal and have norm 1). Also, by redefining if necessary $\ap \to \ap + \frac b2 \overline{w}\cdot v$, $w\to w - (w\cdot v) v$ (which leaves (\ref{eq:eta2}) invariant, upon using (\ref{eq:veta})), we can assume
\begin{equation}\label{eq:wv}
	w \cdot v = 0 \ .
\end{equation}

Now (\ref{eq:eta12}) implies
\begin{equation}\label{scalareq}
|\ap|^2 + |\am|^2 + \b^2 = 1 \ .
\end{equation}
The identity structure is then spanned by $v$, $w$ and 
\begin{equation}
u \equiv \frac{1}{2}\iota_{\overline{w}} \, \omega \ ,
\end{equation}
in terms of which
\begin{equation}
\omega = w \wedge u \ , \qquad -ij = \frac{1}{2}(w \wedge \overline w + u \wedge \overline u) \ .
\end{equation} 
From (\ref{eq:SU2}) we now see that $u$ is also orthogonal to $v$, as well as to $w$ and $\overline w$; moreover, it satisfies $u \cdot \overline u=2$. In other words, 
\begin{equation}
	\{ v, {\rm Re} w, {\rm Im} w, {\rm Re} u, {\rm Im} u\} 
\end{equation}
are a vielbein. 

We can now expand $\psi^1$ and $\psi^2$ in terms of this vielbein. We separate out their even and odd parts:
\begin{equation}\label{spinorparam}
\begin{split}
	\psi^1_+& = \frac{\overline{\ap}}{4}  \exp\left[-ij+\frac{w}{\overline{\ap}}\wedge(\overline{\am} u - b v)\right] \ , \\
	\psi^1_-&= \frac14 (\overline{\ap} v + b w) \wedge \exp\left[-ij+\frac{w}{\overline{\ap}}\wedge(\overline{\am} u - b v)\right] \ ,\\
	\psi^2_+&= -\frac\am 4 \exp\left[-i j + \frac{u}{\am}\wedge(a w - b v)\right]\ , \\
	\psi^2_-&= -\frac14 (\am v + b u) \wedge \exp\left[-i j + \frac{u}{\am}\wedge(a w - b v)\right]\ .	
\end{split}
\end{equation}

\section{Analysis of the conditions for supersymmetry}
\label{generalanalysis}

Having obtained the expansions \eqref{spinorparam} of $\psi^1$, $\psi^2$ in terms of the identity structure on $M_5$,  we can proceed with the study of the system \eqref{susyeq}. In section \ref{geometry} we study the constraints imposed on the geometry of $M_5$ while in section \ref{fluxes} we obtain the expressions of the fluxes in terms of the geometry. The analysis in \ref{geometry} is \emph{local}.

\subsection{Geometry}
\label{geometry}

The equations of the system \eqref{susyeq} which constrain the geometry of $M_5$ are \eqref{susyeqa}, \eqref{susyeqb} and \eqref{susyeqc} with the exception of the three-form part of \eqref{susyeqa} which determines $H$. In the following study of these constraints, it is convenient to introduce the notation
\begin{equation}
\ap \equiv \ap_1 + i \ap_2 \ , \qquad
k_1 \equiv \overline{\ap} v + \b w \ , \qquad
k_2 \equiv - \b v + \ap w \ .
\end{equation}
The zero form part of \eqref{susyeqb}, the one-form part of \eqref{susyeqa}, the two-form part of  \eqref{susyeqc} and the two-form part \eqref{susyeqb}  yield the following set of equations:
\begin{subequations}\label{geomeqs}
\begin{align}
\am & = 0 \ , 
\label{geomeqsam} \\
d\bigl( e^{3A - \phi} \ap_1 \bigr) + 2 e^{2A - \phi} \Im k_1 & = 0 \ ,
\label{geomeqsImk1} \\
d\bigl( e^{4A-\phi}\Re k_1 \bigr) & = 0 \ , 
\label{geomeqsRek1} \\
d\bigl( e^{4A-\phi}\b u \bigr)- 3i e^{3A-\phi} u \wedge k_2 & = 0 \ .
\label{geomeqsu}
\end{align}
\end{subequations}
It can then be shown that the higher-form parts of \eqref{susyeqa}, \eqref{susyeqb} and \eqref{susyeqc} follow from the above equations. 

\eqref{geomeqsam} simplifies quite a bit (\ref{spinorparam}), which now becomes
\begin{equation}\label{spinorparamsimp}
\begin{split}
	\psi^1_+&= \frac14 \overline{\ap} \exp\left[-ij+\frac b{\overline{\ap}} v\wedge w\right] \ ,\qquad
	\psi^2_+= \frac14 (\ap w - b v)\wedge u \wedge e^{-ij} \ , \\
	\psi^1_-&= \frac14 (\overline{\ap} v + b w)\wedge e^{-ij} \ ,\qquad
	\psi^2_-= -\frac14 b u \wedge \exp\left[-ij -\frac{\ap}b v\wedge w\right]\ .	
\end{split}
\end{equation}
It is also interesting to see what the pure spinors $\phi_\pm$ on $M_6$ look like: 
\begin{equation}\label{eq:phiE}
\phi_+ = \frac14 E_1 \wedge E_2 \wedge \exp\left[\frac12 E_3 \wedge \overline{E_3}\right] \ ,
\qquad
\phi_- = E_3 \wedge \exp\left[\frac12(E_1 \wedge \overline{E_1} + E_2 \wedge \overline{E_2})\right] \ ,
\end{equation}
where
\begin{equation}
	E_1 \equiv i e^A b \frac{d \rho}\rho + \ap w - \b v \ ,\qquad
	E_2 \equiv u \ ,\qquad
	E_3 \equiv e^A \overline{\ap} \frac{d \rho}\rho +i (\overline{\ap} v + \b w)\ .
\end{equation}
(\ref{eq:phiE}) are the canonical forms of a type 1 -- type 2 pure spinor pair (where the ``type'' of a pure spinor is the lowest form appearing in it); or, in other words, of a pure spinor pair associated with an SU(2) structure on $M_6$ (although remember that the structure on $M_5$ is the identity). It would be interesting to push this further, and to start an analysis similar to the one in \cite{gabella-gauntlett-palti-sparks-waldram}: in that paper, the language of generalized complex geometry is used to set up a generalized reduction procedure, which eventually leads to a set of four-dimensional equations.

Let us now go back to (\ref{geomeqs}). Given \eqref{geomeqsam}, equation \eqref{scalareq} becomes
\begin{equation}\label{eq:apb}
\ap_1^2 + \ap_2^2 + \b^2 = 1 \ .
\end{equation}
Equations \eqref{geomeqsImk1} and \eqref{geomeqsRek1} can be integrated by introducing local coordinates $y$,
\begin{equation}\label{eq:ya}
y = - \frac{1}{2} e^{3A - \phi} \ap_1 \ ,
\end{equation}
and $x$ such that
\begin{equation}\label{Imk1Rek1dxdy}
\Im k_1 = e^{-2A + \phi} dy \ , \qquad \Re k_1 = e^{-4A + \phi} dx \ .
\end{equation}

$M_5$ possesses an abelian isometry generated by the Killing vector
\begin{equation}
\xi \equiv \frac{1}{2}(\eta_1^\dagger \gamma^m \eta_2 - \eta_2^\dagger \gamma^m \eta_2) \partial_m 
    = -e^A \b (\Re k_2)^\sharp
\end{equation}
where $m = 1, \dots, 5$ and the $\sharp$ superscript denotes the vector dual to the one-form it acts on. A straightforward way to show that $\xi$ is a Killing vector is to work directly with the supersymmetry variations (see appendix \ref{susyvar}) which yield $\nabla_{(m} \, \xi_{\nu)} = 0$ and $\mathcal{L}_\xi \phi = \mathcal{L}_\xi A = 0$, where $\nabla$ is the Levi-Civita connection and $\mathcal{L}_\xi$ is the Lie derivative with respect to $\xi$. It would be interesting to show this directly using the language of generalized complex geometry, and to make contact with the analysis in \cite{gabella-gauntlett-palti-sparks-waldram}. 

Expressing $w,\,v$ in terms of $\Re k_2, \, \Re k_1,$ and $\Im k_1$ we can write the metric on $M_5$ as
\begin{equation}\label{M5metric}
ds^2_{M_5} = ds^2_{\mathcal{C}} + (\Re k_2)^2 + \frac{e^{-4A + 2\phi}}{\b^2}\left[(b^2 + \ap_2^2) e^{-4A} dx^2 + (b^2 + \ap_1^2) dy^2 + 2 \ap_1 \ap_2 e^{-2A} dx dy \right]\ ,
\end{equation}
where $ds^2_{\mathcal{C}} = u \overline{u}$, and $\mathcal{C}$ denotes the two-dimensional subspace spanned by $u$. 

Let us introduce local coordinates $x^I$, $I = 1,2,3$ such that 
\begin{equation}
ds^2_{\mathcal{C}} + (\Re k_2)^2 = g_{IJ}(x^I,x,y) dx^I dx^J \ . 
\end{equation}
$\phi$, $A$ and $\ap_2$ are in principle functions of $x^I$, $x$ and $y$. Given the fact that $\mathcal{L}_\xi \Re k_1 = \mathcal{L}_\xi \Im k_1 = 0$\footnote{Deduced from $\iota_\xi \Re k_1 = \iota_\xi \Im k_1 = 0$ and equation \eqref{Imk1Rek1dxdy}}, we can further introduce a coordinate $x^3 \equiv \psi$ adapted to the the Killing vector 
\begin{equation}
\xi = 3 \partial_\psi \ , 
\end{equation}
in terms of which
\begin{equation}\label{eq:Rek2}
\Re k_2 = - \frac{1}{3} e^A \b D \psi \ ,\qquad D \psi \equiv d\psi + \rho \ , \qquad \rho = \rho_i(x^i,x,y) dx^i \ .
\end{equation}
where $x^i$, $i = 1,2$ are local coordinates on $\mathcal{C}$. Thus 
\begin{equation}
g_{IJ}(x^I,x,y) dx^I dx^J = (g_{\mathcal{C}})_{ij}(x^i,x,y) dx^i dx^j + \frac{1}{9} e^{2A} \b^2  D\psi^2.
\end{equation}
In addition, since $\xi$ is a Killing vector and $\mathcal{L}_\xi \phi = \mathcal{L}_\xi A = 0$, $A$, $\phi$ and $a_2$ are independent of $\psi$.

The exterior derivative on $M_5$ can be decomposed as
\begin{equation}
d = d_2 + d\psi \wedge \partial_\psi + dx \wedge \partial_x + dy \wedge \partial_y \ ,
\end{equation}
where $d_2$ is the exterior derivative on $\mathcal{C}$. We can thus further refine equation \eqref{geomeqsu} as follows:
\begin{subequations}\label{dudecomp}
\begin{align}
d_2 u &= i \rho_0 \wedge u \ , 
\label{dudecomp2} \\
\partial_\psi u &= i u \ , 
\label{dudecomppsi} \\
\partial_x u &= f_1 u \ , 
\label{dudecompx} \\
\partial_y u &= f_2 u \ ,
\label{dudecompy}
\end{align}
\end{subequations}
where 
\begin{subequations}
\begin{align}
\rho_0 &\equiv \rho + *_2 d_2\log\bigl(\b e^{4A - \phi}\bigr) \ , 
\label{rho0} \\
f_1(x^i,x,y) &\equiv - \partial_x \log\bigl(e^{4A - \phi} \b \bigr) + \frac{3 e^{-5A + \phi} \ap_2}{\b^2}  \ , 
\label{fx} \\
f_2(x^i,x,y) &\equiv - \partial_y \log\bigl(e^{4A - \phi} \b \bigr) + \frac{3 e^{-3A + \phi} \ap_1 }{\b^2}\ . 
\label{fy}
\end{align}
\end{subequations}
$*_2$ is the Hodge star defined by $g_\mathcal{C}$, such that $*_2 u = -iu$. 
Integrability of equations \eqref{dudecomp} yields the constraints
\begin{equation}\label{fxfyinteg}
\partial_y f_1 = \partial_x f_2
\end{equation}
and  
\begin{subequations}
\begin{align}
\partial_x \rho_0 
&= - *_2 d_2 f_1 \ , 
\label{dxrho0} \\
\partial_y\rho_0 
&= - *_2 d_2 f_2 \ .
\label{dyrho0}
\end{align}
\end{subequations}

We can write $ds^2_\mathcal{C}$ as
\begin{equation}
ds^2_\mathcal{C} = e^{2\varphi(x^i,x,y)} (dx_1^2 + dx_2^2) \ .
\end{equation}
The Gaussian curvature or one-half the scalar curvature of $\mathcal{C}$, $\ell(x^i,x,y)$, is
\begin{equation}\label{ell}
\ell(x^i, x, y) = - e^{-2\varphi} (\partial_{x_1}^2 + \partial_{x_2}^2) \varphi \ .
\end{equation}

Equations \eqref{dudecomppsi} and \eqref{dudecompx}, \eqref{dudecompy} are solved by 
\begin{equation}
u = e^{\varphi + i\psi} (dx_1 + i dx_2) \ , 
\qquad \partial_x\varphi = f_1 \ , 
\qquad \partial_y\varphi = f_2 \ .
\end{equation}
Equation \eqref{dudecomp2} then yields
\begin{equation}
\rho_0 = \partial_{x_2} \varphi dx_1 - \partial_{x_1} \varphi dx_2 \ ,
\end{equation}
and thus
\begin{equation}\label{d2rho0}
d_2 \rho_0 = \ell(x^i,x,y) \vol_\mathcal{C} \ .
\end{equation}

Compatibility of \eqref{d2rho0} with \eqref{dxrho0}, \eqref{dyrho0} requires that $\ell$ obey the equations
\begin{subequations}\label{eq:d-ell}
\begin{align}
\partial_x \ell + 2 f_1 \ell
&= \Delta_2 f_1 \ ,
\label{dxell} \\
\partial_y \ell + 2 f_2 \ell
&= \Delta_2 f_2 \ ,
\label{dyell}
\end{align}
\end{subequations}
where $\Delta_2 \equiv {d_2}^\dagger d_2 + d_2 {d_2}^\dagger$. The last two equations also follow from \eqref{ell}, bearing in mind that $\Delta_2 \varphi = - e^{-2\varphi} (\partial_{x_1}^2 + \partial_{x_2}^2)\varphi$.

\subsection{Fluxes}
\label{fluxes}

In this section we give the expressions for the fluxes in terms of the geometry of $M_5$. In the following expressions we employ the notation
\begin{equation}
	\begin{split}
	\zeta_1 &\equiv \Re(\ap k_1)^\sharp = - 2y e^A \partial_x - \ap_2 e^{2A - \phi}\partial_y \ ,  \\
	\zeta_2 &\equiv \frac{1}{\b^2} \Im(\ap k_1)^\sharp = \ap_2 e^{4A - \phi} \partial_x - 2y e^{-A}\partial_y \ .
	\end{split}	
\end{equation}

The NSNS three-form flux $H$ is given by the three-form part of equation \eqref{susyeqa}:
\begin{subequations}
\begin{equation}
\begin{split}
	H \ = \  &d\left(\frac{1}{6y} dx \wedge D\psi
	              + \frac{1}{3} e^A \Re (\ap k_1) \wedge D\psi 
	              + \frac{e^{3A - \phi}\ap_2}{2y} \vol_\mathcal{C} \right)  \\
	         &- \frac{1}{6y^2} dx \wedge dy \wedge D\psi 
	         + \frac{e^{-2A}}{y} dx \wedge \vol_\mathcal{C}
	         + \frac{e^{3A-\phi}\ap_2}{2y^2} dy \wedge \vol_\mathcal{C} \ ,	
\end{split}	
\end{equation}
\end{subequations}
where $\Re (\ap k_1) = - 2y e^{-7A+2\phi} dx - \ap_2 e^{-2A + \phi} dy$.

The RR fluxes can be computed from equation \eqref{susyeqd}:
\begin{subequations}\label{F}
\begin{align}
F_0 \ =& \ - 4 e^{2A - 2\phi} \b^2 \partial_y A 
      - e^{-A} \iota_{\zeta_1} d \bigl(e^{A-\phi} \ap_2\bigr) \ ,
\label{F0} \\[.3cm] 
F_2 \ =& \
     \left[
     - 4 e^{-A-\phi} \ap_2  + 4 e^{4A-2\phi} \partial_x A 
     - e^{-5A} \iota_{\zeta_2} d \bigl(e^{5A - \phi} \ap_2\bigr)\right] \vol_\mathcal{C} \nonumber \\
      &+\frac{1}{3} d\bigl(e^{A-\phi}\ap_2\bigr) \wedge D\psi
     + F_0 \frac{1}{3} e^A \Re (\ap k_1) \wedge D\psi  \nonumber \\
      &- \frac{e^{-A}}{\b^2} *_2 d_2\bigl(e^{A - \phi} \ap_2\bigr) \wedge \Im (\ap k_1)
    + 4 e^{-4A} *_2 d_2 A \wedge dx \ ,
\label{F2} \\[.3cm]
F_4 \ =& \ \frac{1}{3} \left[
        e^{-6A} \partial_y\bigl(e^{5A - \phi} \ap_2\bigr) dx
      - e^{-2A} \partial_x\bigl(e^{5A - \phi} \ap_2\bigr) dy
      - 4 e^{-2A} dy
      \right]
      \wedge d\psi \wedge \vol_\mathcal{C} \nonumber \\
      & - \ \frac{1}{3}\left[
        4 e^{-\phi} \ap_2 
       +  e^{-4A} \iota_{\zeta_2} d\bigl(e^{5A - \phi} \ap_2\bigr) 
       \right]
         \Re (\ap k_1) \wedge d\psi \wedge \vol_\mathcal{C} \nonumber \\
      & - \ \frac{1}{3} e^{-10A + 2\phi} *_2 \left[d_2\bigl(e^{5A - \phi} \ap_2\bigr)\right]
        \wedge dx \wedge dy \wedge D\psi \ , 
\label{F4}
\end{align}
\end{subequations}
where $\Im (\ap k_1) = a_2 e^{-4A+\phi} dx - 2y e^{-5A + 2\phi} dy$. 

The fluxes can also be computed from the expression 
\begin{equation}\label{calJfluxes}
F={\cal J}_+ \cdot d_H(e^{-\phi}{\rm Im} \phi_-)
\end{equation}
on $M_6$ \cite{t-reform}. The operator ${\cal J}_+ \cdot$ is associated with the pure spinor $\phi_+$, which can be found in (\ref{eq:phiE}):
\begin{equation}
{\cal J}_+ \cdot = \frac i2 \sum_{i=1}^2 (E_i \wedge \overline{E_i}\llcorner - \overline{E_i}\wedge E_i \llcorner)+\frac i2 (E_3 \llcorner \overline{E_3}\llcorner + E_3 \wedge \overline{E_3}\wedge) \ .
\end{equation}
The degree of difficulty of computing the fluxes from \eqref{calJfluxes} is proportional to the degree of the flux. The opposite is true for computing the fluxes from \eqref{susyeqd}.

\subsection{Bianchi identities}
\label{sub:bianchi}

In order to have a complete supersymmetric AdS$_5 \times M_5$ solution, apart from the conditions for  supersymmetry (which imply the equations of motion \cite{lust-tsimpis}) the Bianchi identities of the fluxes need to be imposed. In this section we study the latter and the extra constraints that follow from their application.

We start with the Bianchi identity of $H$ i.e.\ $dH = 0$. We find that it determines 
\begin{equation}\label{d2rho}
d_2 \rho = e^{-2A}\bigl[6 + 12 y(\partial_y A - f_2) - 6 e^{5A-\phi}  \ap_2 (\partial_x A - f_1)  + 3\partial_x\bigl(e^{5A-\phi}\ap_2\bigr)\bigr] \vol_\mathcal{C} \ .
\end{equation}

Next, we turn to the Bianchi identities of the RR fluxes. The Bianchi identity of $F_0$ just says that it is a constant. The Bianchi identity of $F_2$ is
\begin{equation}
dF_2 - F_0 \, H = 0 \ .
\end{equation}
The non-zero components on the left-hand side are the $dx \wedge \vol_\mathcal{C}$ and $dy \wedge \vol_\mathcal{C}$ components and imposing that they vanish yields the equations:
\begin{subequations}\label{eq:F2bianchi}
	\begin{align}\label{F2Bianchix}
	\partial_x \mathcal{Q} + 2 f_1 \mathcal{Q}
	&- \left[\frac{1}{3}\partial_x\bigl(e^{A-\phi}\ap_2\bigr) - \frac{F_0}{6y}\right] *_2 d_2 \rho - F_0 \frac{e^{-2A}}{y} 
	\\ \nonumber
	&+ \Delta_2\bigl(e^{A - \phi} \ap_2\bigr) \frac{e^{-5A + \phi}\ap_2}{\b^2}
	+ \Delta_2(e^{-4A}) 
	- d_2\bigl(e^{A-\phi} \ap_2\bigr) 
	  \cdot d_2 \left(\frac{2 e^{-5A + \phi} a_2}{\b^2} \right) = 0 \ ,
	\end{align}
	\begin{align}\label{F2Bianchiy}
	\partial_y \mathcal{Q} + 2 f_2 \mathcal{Q}
	&- \frac{1}{3}\partial_y\bigl(e^{A-\phi}\ap_2\bigr) *_2 d_2 \rho - F_0 \frac{e^{3A-\phi}\ap_2}{2y^2} 
	\\ \nonumber
	&- \Delta_2\bigl(e^{A - \phi} \ap_2\bigr) \frac{2 e^{-6A + 2\phi} y}{\b^2} 
	+ d_2\bigl(e^{A-\phi} \ap_2\bigr) 
	  \cdot d_2 \left(\frac{4 e^{-6A + 2\phi} y}{\b^2} \right) = 0 \ . 
	\end{align}
\end{subequations}
where
\begin{equation}
\mathcal{Q}(x^i,x,y) \equiv  - 4 e^{-A-\phi} \ap_2  + 4 e^{4A-2\phi} \partial_x A 
     -  e^{-5A} \iota_{\zeta_2} d \bigl(e^{5A - \phi} \ap_2\bigr) - F_0 \frac{e^{3A - \phi}\ap_2}{2y} 
     \ .
\end{equation}

Finally, the Bianchi identity of $F_4$
\begin{equation}
dF_4 - H \wedge F_2 = 0 \ ,
\end{equation}
is automatically satisfied.

\subsection{Summary so far}
\label{sub:summary}

So far, we have analyzed the constraints imposed by supersymmetry and the Bianchi identities without any Ansatz; let us summarize what we have obtained. 

First of all, we have already determined the local form of the metric: (\ref{M5metric}), (\ref{eq:Rek2}). Most notably, we see the emergence of a Killing vector $\xi$ generating a U(1) isometry, and of a two-dimensional space $\cal C$. The geometry of ${\cal C}$ is constrained by \eqref{dudecomp}. The $S^1$ upon which the U(1) acts is fibered over ${\cal C}$ with $\rho$ being the connection of the fibration. The curvature of the connection is given by \eqref{d2rho}.  

In fact the U(1) isometry is a symmetry of the full solution as it also leaves invariant the fluxes; the latter can be verified by computing the Lie derivative with respect to $\xi$ of the fluxes' expressions as presented in section \ref{fluxes}. This symmetry was to be expected: it is a U(1) R-symmetry corresponding to the R-symmetry of the dual ${\cal N}=1$ field theory. The surface ${\cal C}$ is of less immediate interpretation, but already at this stage it seems to suggest that the field theory should be a compactification on ${\cal C}$ of a six-dimensional field theory. We will see later that this expectation is indeed borne out for the explicit solutions we will find.

We have also reduced the task of finding solutions to a set of partial differential equations on three functions: $a_2$, the dilaton $\phi$, and the warp factor $A$, which in general depend on four variables i.e.\ the coordinates $x^i, x, y$. Supersymmetry equations alone give us (\ref{fxfyinteg}), (\ref{dxell}), (\ref{dyell}). Moreover, the fluxes should satisfy the relevant Bianchi identities, which away from sources give the further equations (\ref{F0}), (\ref{F2Bianchix}), (\ref{F2Bianchiy}). Thus we have a total of six partial differential equations. 
Solving all of them might seem a daunting task, but we will see in the next section that they simplify dramatically with a simple Ansatz. This will allow us to find many explicit solutions.

\section{A compactification Ansatz}
\label{compAnsatz}

We have reduced the general classification problem to a set of six PDEs. To simplify the problem, we will now make an Ansatz.

We assume that $A$, $\phi$ and $a_2$ are functions of $x$ and $y$ only, and that 
\begin{equation}\label{eq:comp-ans}
ds^2_{\mathcal{C}} = e^{2A} ds^2_{\Sigma} (x_1,x_2)  \ .
\end{equation}
In other words, the ten-dimensional metric becomes $ds^2_{10}= e^{2A}(ds^2_{\rm AdS_5} +ds^2_{\Sigma})+ ds^2_{M_3}$. It will soon follow that $\Sigma$ has constant curvature; from now on we will assume it to be a compact Riemann surface $\Sigma_g$. For $g\ge 1$ this involves a quotient by a discrete subgroup, but since no functions depend on its coordinates, this presents no difficulty.

This Ansatz is motivated by the fact that most known solutions in eleven-dimensional supergravity (and hence in massless IIA) are of this type. We also have in mind our original motivation for this paper: finding solutions dual to twisted compactifications of CFT$_6$. If one wants to study a CFT$_6$ on  $\R^4\times \Sigma_g$ rather than on $\R^6$, one needs to replace $ds^2_{\rm AdS_7}= \frac {d \rho^2}{\rho^2} + \rho^2 ds^2_{\R^6}$ with $\frac {d \rho^2}{\rho^2} + \rho^2 (ds^2_{\R^4} + ds^2_{\Sigma})$ in the UV, and then look for a solution that represents the flow to the IR. Our Ansatz is basically that in the IR fixed point this metric is only modified in the $\rho^2$ term multiplying $ds^2_{\Sigma}$, which drops out and becomes a constant.

Whatever its origin, we will now see that this Ansatz is remarkably effective at simplifying the system of PDEs: we will be able to completely classify the resulting solutions. One particular case will be source of many solutions, which will be analyzed in section \ref{sec:sol}. 

\subsection{Simplifying the PDEs} 
\label{sub:pde-comp}

(\ref{eq:comp-ans}) implies
\begin{equation}
f_1 = \partial_x A \ , \qquad f_2 = \partial_y A \ . 
\end{equation}
The integrability condition \eqref{fxfyinteg} is then satisfied trivially, while equations \eqref{dxrho0} and \eqref{dyrho0} yield $\partial_x \rho = \partial_y \rho = 0$ (in the present Ansatz $\rho_0 = \rho$). 

Equations \eqref{d2rho0}, \eqref{d2rho} yield $\ell = e^{-2A} [6 + 3\partial_x (e^{5A - \phi}\ap_2)]$. \eqref{dxell}, \eqref{dyell} are then solved by
\begin{equation}\label{compAnsatzap2}
e^{5A - \phi} \ap_2 = c x + \epsilon \, \qquad c = \mathrm{const.} \ ,
\end{equation}
where $\epsilon= \epsilon(y)$ is a function of $y$ only.
It follows that 
\begin{equation}\label{eq:6+3c}
\ell = e^{-2A} (6 + 3c)
\end{equation}
i.e.\ the Gaussian curvature of $\Sigma_g$ is equal to $6+3c$.

Given the definitions \eqref{fx}, \eqref{fy}, the equations $f_1 = \partial_x A$ and $f_2 = \partial_y A$ become
\begin{subequations}\label{compAnsatzfxfy}
\begin{align}
\partial_x \bigl(e^{10A - 2\phi} b^2 \bigr) &= 6 e^{5A - \phi} \ap_2 \ , \\
\partial_y \bigl(e^{10A - 2\phi} b^2 \bigr) &= 6 e^{7A - \phi} \ap_1 \ . 
\end{align}
\end{subequations}
Recall that $a_1 = -2y e^{-3A+\phi}$ and $b^2 = 1 - a_1^2 - a_2^2$. Using \eqref{compAnsatzap2} we can solve these for 
\begin{subequations}\label{compAnsatzAphi}
\begin{align}\label{eq:compAnsatzphi}
e^{10A-2\phi} - 4y^2 e^{4A} &= c(c+3)x^2 + 2(c+3)\epsilon x + \beta \ ,\\
e^{4A} &= - \frac{\epsilon'}{2y}x - \frac{1}{12y} \left(\beta' - 2 \epsilon\epsilon'\right) \ ,
\label{compAnsatzA}
\end{align}
\end{subequations}
where $\beta= \beta(y)$ is a function of $y$ only, and a prime denotes differentiation with respect to $y$. 

So far we have solved the differential equations imposed by supersymmetry; we now need to impose the Bianchi identities. First, the expression for $F_0$, \eqref{F0}, becomes
\begin{equation}\label{compAnsatzODE3}
e^{12A} F_0 = 
-[c(c + 3)x^2 + 2 (c + 3) \epsilon x  + \beta]   (e^{4A})'
+ e^{4A} \partial_y (c x + \epsilon)^2 + 2 e^{8A} c y\ .
\end{equation}
Recalling  \eqref{compAnsatzA}, we see that this equation is polynomial in $x$, of degree 3. In other words, we can view it as a set of four ODEs in $y$. 

The Bianchi identities for $F_2$, (\ref{eq:F2bianchi}), become
\begin{subequations}
	\begin{align}\label{eq:dexx}
		\del_x^2 ( e^{6A-2 \phi}) &= 0 \ , \\ 
		\label{eq:dexy}
		\del_y\del_x ( e^{6A-2 \phi})  + F_0 \frac{\epsilon'}{2y} &= 0\ .
	\end{align}
\end{subequations}
Substituting equations \eqref{compAnsatzap2} and \eqref{compAnsatzAphi} in  (\ref{eq:dexx}) yields the differential equation
\begin{equation}\label{compAnsatzODE1}
36 (\epsilon')^2 \beta = -(c + 3)(\beta' - 2 \epsilon \epsilon')\left[c \beta' - 2 (c + 6) \epsilon\epsilon'\right] \ .
\end{equation}
Notice that the $x$ dependence has dropped out from this equation. Concerning (\ref{eq:dexy}), just as for (\ref{compAnsatzODE3}), it can be written as a polynomial in $x$ of degree 3, and viewed as four ODEs in $y$. 

So we appear to have reduced the problem to four ODEs from \eqref{compAnsatzODE3}, one from (\ref{eq:dexx}) (which becomes (\ref{compAnsatzODE1})), and four from \eqref{eq:dexy}, for a total of nine ODEs in $y$. However, many of these ODEs actually happen not to be independent from each other. For example, the $x^3$ component of both \eqref{compAnsatzODE3} and (\ref{eq:dexy}) gives
\begin{equation}\label{eq:key}
	4 c(c+3)\left(\frac{\epsilon'}y\right)'+ F_0 \left(\frac{\epsilon'}y\right)^3 = 0 \ ,
\end{equation}
as well as the $x^2$ component of (\ref{eq:dexy}).

To analyze the remaining ODEs, as a warm-up we will first look at the case $F_0=0$, where we will reproduce several known solutions. We will then look at the case $F_0\neq 0$, which we will further split into a generic case where $\epsilon'\neq 0$, and a special case where $\epsilon'=0$; both will give rise to new solutions.


\subsection{\texorpdfstring{$F_0=0$}{F(0) = 0}}
\label{compAnsatzF00}

For $F_0=0$, (\ref{eq:key}) becomes $c(c+3)(\epsilon'-y \epsilon'')=0$. We can then have either  $\epsilon' =y \epsilon''$, $c=-3$, or $c=0$. In the $c=0$ case, actually the $x^2$ coefficient of (\ref{compAnsatzODE3}) gives again $\epsilon' =y \epsilon''$. So this case becomes a subcase of the $\epsilon' =y \epsilon''$ case.

\begin{itemize}
	\item {\bf Case 1:} $\epsilon' =y \epsilon''$. In this case we have
\begin{equation}\label{eq:epsgenF00}
	\epsilon = \frac{1}{2} c_1 y^2 + c_2\ , \qquad c_1, \, c_2 = {\rm const}. \ .
\end{equation}
The $x^3$ component of \eqref{compAnsatzODE3} is (\ref{eq:key}), which we just looked at. The $x^2$ and $x^1$ components both require
\begin{equation}
\left(\frac{\beta'}y\right)'= 2 \frac{c+3}{c+6} c_1 y	 \ .
\end{equation}
The solution to this ODE is
\begin{equation}\label{eq:betagenF00}
	\beta=\frac{c+6}{c+3} \frac{1}{4} c_1^2 y^4 + \frac12 c_3 y^2 + c_4\, \qquad c_3, \, c_4 = {\rm const}. \ .
\end{equation}
The $x^0$ component of \eqref{compAnsatzODE3} then gives 
\begin{equation}\label{eq:c1234}
	(2 c_1 c_2 - c_3)(2(c+6) c_1 c_2 - c c_3 ) + \frac{36}{(c+3)} c_1^2 c_4 = 0 \ .
\end{equation}

Generically this can be solved for $c_4$. In this case, the transformation
\begin{equation}
x \rightarrow x + \frac{\delta}{c}  \ , \qquad 
c_2 \rightarrow c_2 - \delta \ , \qquad
\beta \rightarrow  \beta + \frac{(3 + c) (\delta^2 - 2 \delta \epsilon)}{c}
\end{equation}
leaves the solution invariant and $\delta$ can be chosen such that
\begin{equation}
\beta = \frac{c+6}{c+3} \epsilon^2.
\end{equation}
This branch reproduces the solution obtained from reduction to ten dimensions of the BBBW AdS$_5$ solution of M-theory \cite{bah-beem-bobev-wecht}, as described in appendix \ref{BBBW}. 

This however does not cover the case $c_1=0$. Treating this separately, we find that (\ref{eq:c1234}) leads to $c=0$. This branch reproduces the INST solution \cite{itsios-nunez-sfetsos-thompson}, discussed in appendix \ref{INST}.

\item {\bf Case 2:} $c=-3$. In this case, the $x^2$ component of (\ref{compAnsatzODE3}) gives $\epsilon'=0$. With this, the whole of \eqref{compAnsatzODE3} gives 
\begin{equation}
	2 \beta (\beta'-y \beta'') + y \beta'^2 = 0\ .
\end{equation}
This equation is nonlinear, but if one defines $z=y^2/2$ it becomes $2 \beta \del_z^2 \beta = (\del_z \beta)^2$, which is easily solved by the square of a linear function; in other words, by
\begin{equation}\label{eq:betaMN}
\beta = c_2 (y^2 + 4 c_1)^2 \ , \qquad c_1, \, c_2 = {\rm const}. \ .
\end{equation}
This case reproduces the solution obtained from reduction to ten dimensions of the Maldacena--N\'{u}\~{n}ez AdS$_5$ solution of M-theory \cite{maldacena-nunez}, described in appendix \ref{MN}.

\end{itemize}

\subsection{\texorpdfstring{$F_0 \neq 0$}{F(0) != 0}}

We will divide the analysis in the generic case, where $c\neq 0$ and $-3$, and
two special cases $c=0$ or $-3$. Let us note that from \eqref{compAnsatzODE1}, we see that $\epsilon'=0$ implies either $c=0$ or $-3$; in other words, if $c\neq 0$ and $-3$, then $\epsilon'\neq 0$. On the other hand, from (\ref{eq:key}), we see that $\epsilon' \neq 0$ implies $c\neq 0$ and $-3$; in other words, if $c=0$ or $-3$, then $\epsilon'=0$.

\subsubsection{Generic case} 
\label{ssub:generic}

We begin by analyzing \eqref{compAnsatzODE3} with the aid of  \eqref{compAnsatzODE1} and \eqref{eq:key}. In particular, combining the last two and on the condition that
\begin{equation}\label{betaPrimeCondition}
\beta' \neq \frac{c+3}{c} 2 \epsilon \epsilon' \ ,
\end{equation}
we obtain the following expression for the derivative of 
\eqref{compAnsatzA}:
\begin{equation}
(e^{4A})' = \frac{(\epsilon')^2}{8c(c+3)y^3}\left[F_0 \epsilon' x + \frac{1}{6}F_0(\beta' - 2\epsilon\epsilon') - 4 c y^2\right] \ .
\end{equation}
Substituting  $(e^{4A})'$ and the expression for $\beta$ provided by \eqref{compAnsatzODE1} in \eqref{compAnsatzODE3}, we find that the latter gives
\begin{equation}
\beta' = \frac{c+3}{c} 2 \epsilon \epsilon' \ ;
\end{equation}
this is incompatible with the assumption \eqref{betaPrimeCondition}. 

We thus proceed with the case $c \beta' = (c+3) 2 \epsilon \epsilon'$. Equation \eqref{compAnsatzODE1} fixes 
\begin{equation}
\beta = \frac{c+3}{c} \epsilon^2 \ ,
\end{equation}
while equation \eqref{eq:dexy} follows from \eqref{compAnsatzODE1} and \eqref{eq:key}; the latter can be solved by quadrature. The solution is  
\begin{equation}
	\epsilon = -\frac{2\sqrt{2c(c+3)}}{3F_0^2}(F_0 y - 2c_1)\sqrt{F_0 y + c_1} + c_2 \ , 
	\qquad c_1, \, c_2 = {\rm const}. \ .
\end{equation}
Upon substituting the derived expressions for $\beta$ and $\epsilon$, 
\eqref{compAnsatzODE3} becomes
\begin{equation}
\frac{3y(c+3)(c x + \epsilon)^2}{c(c_1+F_0 y)} = 0 \  ,
\end{equation}
which cannot hold for $c\neq 0$, $-3$.

We conclude that there are no solutions in the generic case.


\subsubsection{Special cases} 
\label{ssub:special}

\begin{itemize}
	\item For $c = 0$, equation \eqref{eq:key} is trivially satisfied, while the $x^1$ component of \eqref{compAnsatzODE3} yields $\epsilon = 0$. Then $\ap_2 = 0$ and this leads to the (unphysical) massive solution of appendix \ref{simpleAnsatz}. 

	\item For $c = -3$, \eqref{eq:key} is again trivially satisfied, while \eqref{compAnsatzODE3} yields the following ODE for $\beta$:
	\begin{equation}\label{eq:ode0}
	e^{12A} F_0 = - \beta \, (e^{4A})' - 6 e^{8A} y \ .
	\end{equation}
	Using (\ref{compAnsatzA}) we see $e^{4A}= -\frac{\beta'}{12y}$. This ODE is nonlinear, and a little tougher than the ones we saw so far in this subsection. Hence we defer its further analysis to the next section. We will see there that it leads to many new AdS$_5$ solutions.
	
\end{itemize}

\section{Compactification solutions} 
\label{sec:sol}

We will now analyze further the case we started considering in section \ref{ssub:special}. We will see that it corresponds to a compactification of the AdS$_7$ solutions considered in \cite{afrt}. Moreover, we will be able to find the most general explicit solution, thus providing a new infinite class of AdS$_5$ solutions.

\subsection{Metric and fluxes} 
\label{sub:metsol}

In section \ref{ssub:special}, we found that there are AdS$_5$ solutions associated with solutions of the ODE (\ref{eq:ode0}). Replacing the expression of $A$ given there, we have
\begin{equation}\label{eq:ode}
\beta \left(y\beta''-\beta'\right) = \frac12 y (\beta')^2 - \frac{F_0}{144 y} (\beta')^3 \ .
\end{equation}
This equation is non-linear; however, it can be rewritten as 
\begin{equation}\label{eq:qtrick}
	(q_5^2)' = \frac29 F_0 \ ,\qquad q_5 \equiv - \frac{4 y \sqrt{\beta}}{\beta'} \ .
\end{equation} 
We will see later that $q_5$ has actually a useful physical interpretation (similar to the $q$ of \cite{afrt}): it will turn out to be related to D8-brane positions. In any case, the trick (\ref{eq:qtrick}) allows us to solve the ODE (\ref{eq:ode}): indeed we can write $16 y^2 \frac{\beta}{(\beta')^2}= \frac29 F_0 (y-\hat y_0)$, which can now be integrated by quadrature. 

We will postpone the detailed analysis of the solutions of (\ref{eq:ode}) to sections \ref{sub:nice} and \ref{sub:gen}. For the time being, in this subsection we will collect various features of the resulting AdS$_5$ solutions.

The internal metric for the class we are considering can be extracted from the general expression (\ref{M5metric}). However, at first its global meaning is not transparent. It proves useful to trade the coordinate $x$ for a new coordinate $\theta$, defined by 
\begin{equation}
	\cos\theta = \frac{-3x+ \epsilon}{\sqrt{\beta}}\ .
\end{equation}
The metric then becomes
\begin{equation}\label{eq:metsol}
	ds^2_{M_5}= e^{2A}ds^2_{\Sigma_g} + ds^2_{M_3} \ ,\qquad
	ds^2_{M_3} = dr^2 + \frac19 e^{2A}(1-\ap_1^2)ds^2_{S^2}\ .
\end{equation}
Here 
\begin{equation}\label{eq:S2}
	ds^2_{S^2}= d\theta^2 + \sin^2\theta D\psi^2
\end{equation}
is the metric of the round $S^2$, fibered over $\Sigma_g$, which is a Riemann surface of Gaussian curvature $-3$ (recalling (\ref{eq:6+3c}), and $c=-3$) and hence $g\ge 2$; The new coordinate $r$ is defined by 
\begin{equation}\label{eq:dr}
	dr = \frac{e^{3A}}{\sqrt{\beta}}dy\ .
\end{equation}
Moreover, from (\ref{eq:ya}) and (\ref{compAnsatzAphi}) we have
\begin{equation}\label{eq:Aphisol}
	1-a_1^2= \frac{3 \beta}{3 \beta- y \beta'}\ ,\qquad e^{4A}=-\frac{\beta'}{12y} \ ,\qquad e^{\phi} = \frac{\sqrt{3} e^{5A}}{\sqrt{ 3\beta - y\beta'}}\ .
\end{equation}
We can now remark that the $q_5$ defined in (\ref{eq:qtrick}) is 
\begin{equation}\label{eq:q5}
	q_5 \equiv e^{-\phi} R_{S^2} \equiv \frac13 e^{A-\phi} \sqrt{1- \ap_1^2}= - \frac{4 y \sqrt{\beta}}{\beta'}\ .
\end{equation}
$R_{S^2}= \frac13 e^A  \sqrt{1- \ap_1^2} $ is the radius of the round $S^2$, as inferred from (\ref{eq:metsol}). The role of this particular combination of the radius and dilaton will become clearer in section \ref{sub:fluxq}. 

From (\ref{eq:q5}) and (\ref{eq:Aphisol}) we see that for the solution to make sense we must require
\begin{equation}\label{eq:ineq}
	\beta \ge 0 \ ,\qquad - \frac{\beta'}y \ge 0 \ .
\end{equation}

We can now also obtain the fluxes, from the formulas in section \ref{fluxes}. We have
\begin{equation}\label{eq:F2sol}
	F_2 = q_5 \left[- ({\rm vol}_{S^2} + 3 \cos\theta {\rm vol}_{\Sigma_g})+ \frac13 F_0 \ap_1 e^{A+\phi} {\rm vol}_{S^2} \right]\ ,
\end{equation}
where ${\rm vol}_{S^2}\equiv \sin\theta d\theta \wedge D\psi$. The four-form flux reads
\begin{equation}\label{eq:F4sol}
	F_4 = \frac13 {\rm vol}_{\Sigma_g} \wedge \left[\frac{2 y \beta}{3 \beta - y \beta'} \cos\theta {\rm vol}_{S^2} + \sin^2\theta D\psi\wedge dy\right]\ .
\end{equation}
When $F_0\neq 0$, we need not give an expression for $H$: as usual for massive IIA, it can be written as $H=dB$, where 
\begin{equation}\label{eq:Bb}
	B= \frac{F_2}{F_0}+ b\ ,
\end{equation}
where $b$ is a closed two-form. When $F_0=0$, the only solution in the class we are considering in this section is the Maldacena--N\'u\~nez solution; an expression for $B$ is presented for that case in (\ref{eq:Bmn}).

We can observe already now that the metric (\ref{eq:metsol}) and the flux (\ref{eq:F2sol}) look related to those for the AdS$_7$ solutions in \cite{afrt}; see (4.16) and (4.9) there. The expressions are very similar; one obvious difference is that the three-dimensional metric in (\ref{eq:metsol}) is fibered over $\Sigma_g$, and that the flux (\ref{eq:F2sol}) has extra legs along $\Sigma_g$. Except for a few numerical factors, everything seems to correspond nicely; the role of $x$ in \cite{afrt} seems to be played here by $a_1$:
\begin{equation}
	x\ {\rm in\  \cite{afrt}} \to a_1\ {\rm here}.
\end{equation} 
Actually this correspondence can be justified a little better. In \cite{afrt}, $x$ is the zero-form part of ${\rm Im} \psi^1_+$ (in that paper's notation), which is the calibration for a D6-brane extended along AdS$_7$. The analogue of this in our case would be a D6-brane extended along AdS$_5\times \Sigma_g$; the relevant calibration is the part along $u\wedge \bar u$ of ${\rm Im}  \psi^1_+$ of the present paper. Looking at (\ref{spinorparam}), we see that that is indeed ${\rm Re} \ap= \ap_1$.

Motivated by this, in this section we will also use the name
\begin{equation}\label{eq:x5}
	x_5 \equiv a_1\ .
\end{equation}
This $x_5$ is meant to evoke the $x$ in \cite{afrt}, and is not to be confused with the coordinate $x$ we temporarily used in sections \ref{generalanalysis} and \ref{compAnsatz}.


\subsection{\texorpdfstring{Correspondence with AdS$_7$}{Correspondence with AdS(7)}} 
\label{sub:ads7}

We will now show that solutions of the type considered in section \ref{ssub:special} are in one-to-one correspondence with the AdS$_7$ solutions in \cite{afrt}. The map we will find is directly inspired from a similar map from AdS$_4$ to AdS$_7$ found in \cite{rota-t}. It would be possible to present our new AdS$_5$ solutions perfectly independently from the map to AdS$_7$; in fact, in finding the analytic solutions the map does not help at all. However, the existence of the map tells us right away that infinitely many regular solutions do exist, and what data they depend on. 

Let us start from (\ref{eq:ode}). Using the definition (\ref{eq:dr}), the expressions (\ref{eq:Aphisol}) and the expression $x_5= a_1 = -2y e^{-3A+\phi}$ from (\ref{eq:x5}), (\ref{eq:ya}), we can see that
\begin{align}\label{eq:oder5}
	&\del_r \phi = \frac14 \frac{e^{-A}}{\sqrt{1-x_5^2}} (11 x_5 -2 x_5^3 + (2x_5^2-5)F_0 e^{A+\phi}) \ , \nonumber\\
	&\del_r x_5 = -\frac12 e^{-A}\sqrt{1-x_5^2} (4-x_5^2+x_5 F_0 e^{A+\phi}) \ ,\\
	&\del_r A = \frac14 \frac{e^{-A}}{\sqrt{1-x_5^2}} (3x_5 - F_0 e^{A+\phi}) \nonumber\ .
\end{align}
Conversely, given a solution to this system, one may define $\beta = e^{10 A-2 \phi}(1-x_5^2)$, $y= -\frac12 x_5 e^{3A-\phi}$ (with an eye to \eqref{compAnsatzAphi}, (\ref{eq:ya}), which correspond to (\ref{eq:Aphisol})); if one then eliminates $r$ from (\ref{eq:oder5}), the resulting equations imply $\beta'= -12 y e^{4A}$ (the second in (\ref{eq:Aphisol})), and (\ref{eq:ode}). So the system (\ref{eq:oder5}) is in fact an equivalent way to characterize our solutions. It looks much more complicated than the original ODE (\ref{eq:ode}). We write it because it bears an uncanny resemblance with the system in \cite[Eq.(4.17)]{afrt}: a few numerical factors have changed, and two new terms have appeared. This suggests that there might be a close relationship between solutions of one system and solutions of the other. This is in fact the case: to any solution $(\phi_5,x_5,A_5)$ of (\ref{eq:oder5}) one can associate a solution $(\phi_7,x_7,A_7)$ of  \cite[Eq.~(4.17)]{afrt} given by
\begin{equation}\label{eq:map57}
\begin{split}
	&e^{\phi_7} = \left(\frac 34\right)^{1/4} \frac{e^{\phi_5}}{\sqrt{1-\frac14 x_5^2}} \ ,\qquad
	e^{A_7} = \left(\frac 43\right)^{3/4} e^{A_5}\ ,\\
	&x_7 = \left(\frac 34\right)^{1/2}\frac{x_5}{\sqrt{1-\frac14 x_5^2}} 
	\ ,\qquad r_7 =  \left(\frac 43\right)^{1/4} r_5\ .
\end{split}	
\end{equation}
Comparing (\ref{eq:metsol}) with \cite[Eq.~(4.16)]{afrt}, we find that the map acts on the metrics as
\begin{equation}\label{eq:57metric}
\begin{split}
	 e^{2A_5}(ds^2_{{\rm AdS}_5}+ ds^2_{\Sigma_g}) + dr^2_5 + &\frac{1-x_5^2}9 e^{2A_5} ds^2_{S^2} \to 
	 \\ 
	&\sqrt{\frac43}\left(\frac43 e^{2A_5} ds^2_{{\rm AdS}_7} + dr^2_5 +\frac{e^{2A_5}}{12}\frac{1-x_5^2}{1-\frac14 x_5^2}ds^2_{S^2}\right) \ . 
\end{split}
\end{equation} 
Conversely, to any solution $(\phi_7, x_7, A_7)$ of \cite[Eq.(4.17)]{afrt}, one can associate a solution $(\phi_5, x_5, A_5)$ of (\ref{eq:oder5}) given by 
\begin{equation}\label{eq:map75}
\begin{split}
	&e^{\phi_5} = \left(\frac 43\right)^{1/4} \frac{e^{\phi_7}}{\sqrt{1+\frac13 x_7^2}} \ ,\qquad
	e^{A_5} = \left(\frac 34\right)^{3/4} e^{A_7}\ ,\\
	&x_5 = \left(\frac 43\right)^{1/2}\frac{x_7}{\sqrt{1+\frac13 x_7^2}} 
	\ ,\qquad r_5 =  \left(\frac 34\right)^{1/4} r_7\ .
\end{split}
\end{equation}
This inverse map now acts on the metrics as
\begin{equation}\label{eq:75metric}
\begin{split}
	e^{2A_7} ds^2_{{\rm AdS}_7} + dr_7^2 + &\frac{1-x_7^2}{16} e^{2A_7} ds^2_{S^2} \ \to \ \\
	&\sqrt{\frac34}\left(\frac34e^{2A_7}(ds^2_{{\rm AdS}_5}+ ds^2_{\Sigma_g}) +dr_7^2 + \frac1{12} \frac{1-x_7^2}{1+\frac13 x_7^2}e^{2A_7} ds^2_{S^2}\right)\ .
\end{split}
\end{equation}
The simplicity of this map is basically a generalization of the simple Maldacena--N\'u\~nez solution \cite{maldacena-nunez}, with the $1+\frac13 x_7^2$ factor ultimately playing the role of the $\Delta=1+\sin^2\theta$ factor in \cite{maldacena-nunez}.

One can also apply (\ref{eq:map57}) directly to (\ref{eq:Aphisol}), and infer the expressions for the variables of the seven-dimensional solution:
\begin{equation}
	e^{A_7}= \frac23 \left(-\frac {\beta'}y\right)^{1/4}\ ,\qquad
	x_7 = \sqrt{\frac{-y \beta'}{4 \beta - y \beta'}} \ ,\qquad e^{\phi_7}=\frac{(-\beta'/y)^{5/4}}{12\sqrt{4 \beta - y \beta'}}\ .
\end{equation}
Moreover, $dr_7=\left(\frac34\right)^2 \frac{e^{3A_7}}{\sqrt{\beta}}dy$. 

In \cite{afrt}, solving the system of ODEs in \cite[Eq.(4.17)]{afrt} was only part of the problem. First, one had to take care of flux quantization; second, most solutions include D8's, and one must take care that supersymmetry be preserved also on top of them. We will see in section \ref{sub:fluxq} that the relevant conditions also map nicely under (\ref{eq:map57}); that will lead us to conclude that there are infinitely many AdS$_5$ solutions, each one of them corresponding to the AdS$_7$ solutions in \cite{afrt,gaiotto-t-6d}. Moreover, the map is quite simple: for example, it acts on the metrics as in (\ref{eq:75metric}). 


\subsection{Regularity analysis} 
\label{sub:reg}

We showed that solutions of (\ref{eq:ode}) are in one-to-one correspondence with solutions of the system of ODEs relevant for AdS$_7$ solutions. However, (\ref{eq:ode}) looks much simpler than that system; hence one may hope to learn more about both the AdS$_5$ and the AdS$_7$ solutions by studying it.

In this subsection we will see what boundary conditions on (\ref{eq:ode}) have to be imposed in order to obtain compact and regular solutions. 

We saw in (\ref{eq:metsol}) that the internal metric consists of an $M_3$ fibered over a Riemann surface $\Sigma_g$; $M_3$ is itself a fibration of $S^2$ over a one-dimensional space with coordinate $r$.

To make $M_3$ compact, we can use the same logic as for the AdS$_7$ solutions of \cite{afrt}. One might think of making it compact by periodically identifying $r$, but this doesn't work for the same reason as in \cite[Eq.~(4.24)]{afrt}: the quantity $y= -\frac12 e^{3A-\phi}x_5$ is monotonic --- from (\ref{eq:oder5}) we see $\del_r y = e^{2A-\phi}\sqrt{1-x_5^2}$, which is always positive; or also, directly from (\ref{eq:dr}) we see $\frac{\del y}{\del r}= e^{-3A}\sqrt{\beta}$. So periodically identifying $r$ is not an option. The other way to make $M_3$ compact is to make the $S^2$ shrink for two values of $r$, just like in \cite{afrt}. This is what we will now devote ourselves to.

To make the $S^2$ shrink, we should make the coefficient $(1-a_1^2)$ in (\ref{eq:metsol}) go to zero, which, recalling (\ref{eq:Aphisol}), can be accomplished by making $\beta$ vanish. If $\beta$ has a single zero, 
\begin{equation}\label{eq:single0}
	\beta = \beta_1 (y-y_0) + O(y-y_0)^2 \ ,\qquad
\end{equation} 
the metric (\ref{eq:metsol}) near $y_0$ is proportional to 
\begin{equation}
	\frac{dy^2}{4 (y-y_0)} + (y -y_0) ds^2_{S^2} \ ,
\end{equation}
which in fact upon defining $r=\sqrt{y-y_0}$ turns into 
\begin{equation}
	dr^2 + r^2 ds^2_{S^2}\ ,
\end{equation}
which is the flat metric on $\R^3$. Hence if $\beta$ has a single zero at $y_0\neq 0$ the metric is regular. 

One might wonder what happens if $\beta$ has a double zero:
\begin{equation}\label{eq:double0}
	\beta = \beta_2 (y-y_0)^2 + O(y-y_0)^3 \ . 
\end{equation}
In this case,  (\ref{eq:metsol}) is proportional to $\frac{dy^2}{\sqrt{y-y_0}}+ (y-y_0)^{3/2}ds^2_{S^2}$, which upon defining $\rho=y-y_0$ turns into
\begin{equation}
	\frac1{\sqrt{\rho}} ( d \rho^2 + \rho^2 ds^2_{S^2})\ ; 
\end{equation}
we also have $e^A\sim \rho^{1/4}$, $e^\phi \sim \rho^{3/4}$. This is obviously not a regular point, but it is the local behavior appropriate for a D6 stack whose transverse directions are $\rho$ and the $S^2$. 

Higher-order zeros do not lead to anything of physical relevance, and in fact they would not lead to solutions, as we will see later. However, given that we have obtained boundary conditions corresponding to a regular point and to presence of a D6 stack, it is natural to wonder whether we can find boundary conditions corresponding to presence of an O6. This is realized when 
\begin{equation}\label{eq:sqrt}
	\beta  = \beta_0 + \beta_{1/2} \sqrt{y-y_0} + O(y-y_0)\ ;
\end{equation}
in this case the metric is proportional to $(y-y_0)^{1/4}\left(\frac{d y^2}{y-y_0}+ 16 \alpha_0^2 ds^2_{S^2}\right)$, with $\alpha_0\equiv \frac{\beta_{1/2}}{\beta_0}$. With the definition $\rho=\sqrt{y-y_0}$, this turns into
\begin{equation}
	\sqrt{\rho}\left(d \rho^2 + 4 \alpha_0^2 ds^2_{S^2}\right)\ ;
\end{equation}
moreover, $e^A\sim \rho^{-1/4}$, $e^{\phi} \sim \rho^{-3/4}$. These are the appropriate behaviors for fields near the beginning of an O6 hole: to see this, one can start from the flat space O6 metric, given by $ds^2_\perp = H^{1/2}(d \rho^2 + \rho^2 ds^2_{S^2})$, $e^A= H^{1/4}$, $e^\phi \propto H^{3/4}$, $H= 1-\frac{\rho_0}\rho$, and expand around $\rho=\rho_0$, which is indeed the boundary of the O6 hole.

This concludes our study of the physically relevant boundary conditions for the ODE (\ref{eq:ode}); as it will turn out, these are the only ones which are actually realized in its solutions. Later in this section we will turn to the task of finding such solutions. 


\subsection{Flux quantization, D8 branes} 
\label{sub:fluxq}

Before we look at explicit solutions, we will discuss flux quantization. We will also introduce D8-branes in our construction, as was done in \cite{afrt}. This subsection is in many ways similar to \cite[Sec.~4.8]{afrt}, which the reader may want to consult for more details.

We will start with some preliminary comments about the $B$ field. In (\ref{eq:Bb}) we expressed it in terms of a closed two-form $b$. We will need this second term because the term $\frac{F_2}{F_0}$ in (\ref{eq:Bb}) will jump as we cross a D8 (since $F_0$ will jump there, by definition). More precisely, looking at $F_2$ we see that only the term proportional to ${\rm vol}_{S^2} + 3 \cos\theta {\rm vol}_{\Sigma_g}$ jumps (since in the other term an $F_0$ cancels out). Thus we can limit ourselves to considering $b$ of the form 
\begin{equation}
	b = b_0 ({\rm vol}_{S^2} + 3 \cos\theta {\rm vol}_{\Sigma_g})\ ,
\end{equation} 
which is indeed closed (while ${\rm vol}_{S^2}=\sin\theta d\theta \wedge D\psi$ would not be, because of the presence of $\rho$). (\ref{eq:Bb}) now becomes
\begin{equation}\label{eq:Bext}
	B = \left(b_0-\frac{q_5}{F_0}\right)({\rm vol}_{S^2} + 3 \cos\theta {\rm vol}_{\Sigma_g}) + \frac{q_5}3 x_5 e^{A+\phi} {\rm vol}_{S^2}\ .
\end{equation}
At the poles, for regularity we should have that what multiplies ${\rm vol}_{S^2}$ should go to zero. 

However, more precisely $B$ should be understood as a ``connection on a gerbe''. Concretely, this means that it is not necessarily a globally well-defined two-form. On a chart intersection $U \cap U'$, $B_U-B_{U'}$ can be any closed two-form whose periods are integer multiples of $4\pi^2$ (known as a ``large gauge transformation''). This translates into the requirement that the coefficient of ${\rm vol}_{S^2}$ in (\ref{eq:Bext}) should wind $\pi\times$ an integer number of times in going from the north to the south pole. Alternatively, using Stokes' theorem, we see that the integral of $H$ between $r_{\rm N}$ and $r_{\rm S}$ (the positions of the two poles) is 
\begin{equation}
\int_{M_3}H = \int_{S^2}\int_{r_{\rm N}}^{r_{\rm S}} dr H = \int_{S^2} (B(r_{\rm N})-B(r_{\rm S})) \ ; 
\end{equation}
thus $\int_{M_3} H$ will be an integer multiple of $4\pi^2$, in agreement with flux quantization.

After these comments on the NSNS flux $H$, let us now consider the RR fluxes. First of all, the zero-form should satisfy $F_0= \frac{n_0}{2\pi}$, $n_0 \in \zz$. For the higher forms, we should consider 
\begin{equation}
	\tilde F_2 \equiv F_2 - B F_0 \ ,\qquad \tilde F_4 \equiv F_4 - B \wedge F_2 + \frac12 B\wedge B F_0 \ ,
\end{equation}
which are $d$-closed (unlike the original $F_2$ and $F_4$, which in our notation are $(d-H\wedge)$-closed). Flux quantization imposes that those should have integer periods. For the two-form we simply have
\begin{equation}
	\tilde F_2 = - b F_0 = - b_0 F_0 ({\rm vol}_{S^2} + 3 \cos\theta {\rm vol}_{\Sigma_g})\ .
\end{equation}
Integrating this on the fiber $S^2$ and imposing that it is of the form $2\pi n_2$, $n_2\in \zz$, we find
\begin{equation}\label{eq:b0}
	b_0 = -\frac{n_2}{2 F_0}=-\pi\frac{n_2}{n_0}\ ,
\end{equation}
just like in \cite{afrt}. A gauge transformation will change $b_0\to b_0 + k\pi$, and simultaneously $n_2\to n_2 - k$, so that (\ref{eq:b0}) remains satisfied. 

Near the north and south pole it is convenient to work in a gauge where $B$ is regular; then $\int \tilde F_2 \to \int F_2$, and $n_2$  is determined by setting to zero the limit near the pole of $\left(b_0-\frac{q_5}{F_0} + \frac{q_5}3 x_5 e^{A+\phi}\right)$, the coefficient of ${\rm vol}_{S^2}$ in (\ref{eq:Bext}). For a regular point, $n_2$ near the pole is zero, and both $q_5\to 0$ and $q_5 x_5 e^{A+\phi}\to 0$. For a stack of $n_2$ D6-branes, $q_5 x_5 e^{A+ \phi}\to 0$, and $q_5\to -\frac{n_2}2$. In section \ref{sub:reg}, we saw that presence of a D6 corresponds to a double zero in $\beta$, (\ref{eq:double0}). The condition we just saw will then discretize the parameter $\beta_2$, giving
\begin{equation}\label{eq:b2O6}
	\beta_2 = \left(\frac{4 y_0}{n_2}\right)^2\ .
\end{equation}

An O6 point is different: $n_2=\pm1$ (depending on whether we are considering the north or south pole), $q_5\to 0$, but $\frac{q_5}3 x_5 e^{A+\phi}$ is non zero, and will have to tend to $-\frac{n_2}{2 F_0}$. Again in section \ref{sub:reg} we saw that an O6 corresponds in our class of solutions to the presence of a square root, (\ref{eq:sqrt}). Flux quantization will then fix 
\begin{equation}
	\beta_0 = \left(\frac{18 y_0}{F_0}\right)^2 \ .
\end{equation}

The four-form $\tilde F_4$ can now be written, after some manipulations, as
\begin{equation}\label{eq:F4t}
	\tilde F_4 = \left(\frac3{F_0}\left(-q_5^2 +\frac{n_2^2}4 \right) \cos \theta {\rm vol}_{S^2} + \frac13 \sin^2 \theta D\psi \wedge dy \right) \wedge {\rm vol}_{\Sigma_g}\ .
\end{equation}
Using (\ref{eq:qtrick}) we can also write $\tilde F_4 = d\tilde C_3$, where 
\begin{equation}\label{eq:C3}
	\tilde C_3= \frac3{2F_0}\left(-q_5^2 + \frac{n_2^2}4\right)\sin^2\theta D\psi \wedge {\rm vol}_{\Sigma_g} \ .
\end{equation}
If both poles are regular points, $\tilde C_3$ is a regular form. Indeed, as we saw, at such a pole we should have $n_2=0$ and $q_5\to 0$. So the coefficient $\left(-q_5^2 + \frac{n_2^2}4\right)$ will actually go to zero at the pole. Now, using the fact that $\beta$ has a single zero (\ref{eq:single0}), from (\ref{eq:dr}) and (\ref{eq:q5}) we see that $q_5$ starts with a linear power in $r$. Hence we have
\begin{equation}
	\tilde C_3 \sim r^2 \sin^2\theta D\psi \wedge {\rm vol}_{\Sigma_g}\ .
\end{equation} 
Now, $r^2\sin^2\theta d\psi$, going from spherical to cartesian coordinates $x^i$, $i=1,2,3$, is proportional to $x^1 dx^2 - x^2 dx^1$, and hence is regular. All in all, we conclude that $\tilde F_4$ does not have any non-zero periods, since it is exact. In presence of a D6 or O6 point, it is best to go back to (\ref{eq:F4t}). The space is topologically an $S^3$ fibration over $\Sigma_g$; standard topological arguments tell us that its cohomology is just the product of that of $S^3$ and that of $\Sigma_g$. As such it would have no four-cycles. Thus so far flux quantization for $\tilde F_4$ is not an issue.

We will now introduce D8-branes. We will consider them to be extended along all directions except $r$. Their treatment is very similar to \cite{afrt} and we will be brief. The defining feature of a D8 stack is that the Romans mass $F_0$ jumps as we go across them. Let us call $n_0$ and $n_0'$ the flux integers on the two sides. Moreover, we will allow the D8's to have non-zero worldsheet flux, which can also be thought of as a smeared D6 charge. This will make the flux integer for $\tilde F_2$ jump as well; we will call $n_2$ and $n_2'$ its value on the two sides. The ``slope'' $\mu\equiv\frac{\Delta n_2}{\Delta n_0}\equiv \frac{n_2'-n_2}{n_0'-n_0}$ needs to be an integer. With this notation, imposing that (\ref{eq:Bext}) be continuous we find the condition
\begin{equation}\label{eq:qD8}
	q_5|_{\rm D8} = \frac12 \frac{n_2' n_0 - n_2 n_0'}{n_0'-n_0} = \frac12 (-n_2 + \mu n_0) = \frac12 (-n_2'+\mu n_0')\ .
\end{equation}
This is to be read as a condition fixing the D8's position. 

One might now also wonder whether the flux of $\tilde F_4$ along $\Sigma_g\times S^2$ might jump between D8's, as does the integral of $\tilde F_2$. But actually $\int_{S^2}\cos \theta \vol_{\rm S^2}=0$. So even in presence of D8's we need not worry about flux quantization for $\tilde F_4$.

Crucially, (\ref{eq:qD8}) is exactly the same condition that was found for D8-branes in \cite[Eq.~(4.45)]{afrt}. The function called $q$ in that paper, which we will call $q_7$ here, is not exactly the same as our $q_5$ defined in (\ref{eq:q5}): indeed $q_7\equiv \frac14 e^{A_7-\phi_7}\sqrt{1-x_7^2}$. However, using the map (\ref{eq:map57}), we see that the different overall factor is reabsorbed:\footnote{Actually, the condition that the system (\ref{eq:oder5}) be mapped to the similar system \cite[Eq.~(4.17)]{afrt} for AdS$_7$ solutions only fixed the map (\ref{eq:map57}) up to a constant. We fixed the constant so that (\ref{eq:qD8}) would look exactly equal to \cite[Eq.~(4.45)]{afrt}.} 
\begin{equation}
	q_5 = q_7\ .
\end{equation} 
So (\ref{eq:qD8}) fixes the D8's at exactly the same position in an AdS$_5$ solution and in its AdS$_7$ solution. 

Since (\ref{eq:qD8}) was found by imposing that $B$ should be continuous, it looks easy to impose the condition on flux quantization. As remarked earlier, by Stokes' theorem we can relate the integrality of $H$ to the periodicity of the coefficient of ${\rm vol}_{S^2}$ in $B$. (This periodicity was expressed visually in several figures in \cite{afrt,gaiotto-t-6d}, as a dashed green line.) However, in presence of D8's one might encounter a region where $F_0=0$; generically such a region will exist (although there are also ``limiting cases'' where it does not exist; see \cite[Sec.~4.2]{gaiotto-t-6d}). In such a region, (\ref{eq:Bb}) (and hence (\ref{eq:Bext})) cannot be used; we have to resort to (\ref{eq:Bmn}). This allows to write a general expression for the integral of $H$, as shown in \cite[Eq.(4.7)]{gaiotto-t-6d}. 

Since we are going to simplify that formula for AdS$_7$ solutions, let us review it quickly here. To simplify things a bit, one derives first an expression for the integral in the ``northern hemisphere'', between $x_7=1$ and $x_7=0$; it can be shown that $x_7=0$ is in the massless region, where $F_0=0$. There might be many D8's; let D8$_n$ be the one right before the massless region, $\{n_{2,n},n_{0,n}=0\}$ the flux parameters right after it, and $\{n_{2,n-1},n_{0,n-1}\}$ the ones right before it. Then we can divide the integral into a contribution from the massive region and one from the massless region: 
\begin{equation}\label{eq:Hnorth}
\begin{split}
	\int_{\rm north} H &= \int_{r_{\rm N}}^{{\rm D8}_n} H + \int_{{\rm D8}_n}^{x=0} H \\
	 &= 4\pi\left[ q_7 \left(\frac{x_7}4  e^{A_7+\phi_7}-\frac1{F_{0,n-1}}\right)-\frac{n_{2,n-1}}{2F_{0,n-1}}+\frac3{32}\frac{R^3}{n_{2,n}}\left( x_7 - \frac{x_7^3}3\right)\right]_{{\rm D8}_n}\\
	&=4 \pi \left[-\pi \mu_n + \frac14 q x_7 e^{A_7+\phi_7} -\frac14 q_7 x_7 e^{A_7+\phi_7}\frac{3-x_7^2}{1-x_7^2}\right]_{{\rm D8}_n}\\
	&=4\pi\left[-\pi \mu_n + \frac{R^3}{16 n_{2,n}}x_7\right]_{{\rm D8}_n}\ .
\end{split}
\end{equation}
We have used that for the massless solution $-8 q_7 \frac{e^{A_7+\phi_7}}{1-x_7^2}= -2 \frac{e^{2A_7}}{\sqrt{1-x_7^2}}=\frac{R^3}{n_2}$, where $R$ is a constant. After this simplification, and putting together the contribution from $\int_{\rm south} H$ from the ``southern hemisphere'', we can write
\begin{equation}\label{eq:Hfluxq}
	N \equiv -\frac1{4\pi^2}\int H = (|\mu_n| + |\mu_{n+1}|) + \frac1{4\pi} e^{2A(x=0)}(|x_n|+|x_{n+1}|)\ ,
\end{equation}
where $x_n$ and $x_{n+1}$ are the values of $x_7$ at the branes D8$_n$ and D8$_{n+1}$.\footnote{The $\mu_i$ and $x_i$ before the massless region are positive, while those after the massless region are negative.} 

To derive a similar expression for AdS$_5$ solutions, we follow a similar logic. It proves convenient to use from the very beginning $(A_7,x_7,\phi_7)$ variables, which are related to $(A_5,x_5, \phi_5)$ variables via (\ref{eq:map57}). We can use (\ref{eq:Bext}) and (\ref{eq:Bmn}), the latter of which is already expressed in terms of $x_7$. Some factors in the computation change, but remarkably the result turns out to be exactly the same as in (\ref{eq:Hfluxq}). As a consequence, if the $H$ flux quantization is satisfied for an AdS$_7$ solution, it is also satisfied for an AdS$_5$ solution, and viceversa.

So the conclusion of this section is that the flux quantization conditions and the constraints fixing the D8-brane positions are all precisely mapped by (\ref{eq:map57}), in such a way that if they are satisfied for an AdS$_7$ solution they are also automatically satisfied for an AdS$_5$ solution. This proves that the map (\ref{eq:map57}) produces infinitely many AdS$_5$ solutions.


\subsection{The simplest massive solution} 
\label{sub:nice}

We will now start studying solutions to (\ref{eq:ode}), and their associated physics. We have already indicated in (\ref{eq:qtrick}) how to solve it analytically. However, in this section we will warm up by a perturbative study, which we find instructive and which will allow us to isolate a particularly nice and useful solution. 

In section \ref{sub:reg} we studied the boundary conditions for the ODE (\ref{eq:ode}). We can now proceed to study it in the neighborhood of such a solution. We will do so by assuming analytic behavior around $y_0$: $\beta =\sum_{k=1}^\infty \beta_k (y-y_0)^k$, by plugging this Taylor expansion in (\ref{eq:ode}), and solving order by order. 

Already at order zero we find
\begin{equation}\label{eq:branch}
 	\left(\beta_1- \frac{72 y_0^2}{F_0}\right)\beta_1^2= 0 \ .
\end{equation}
The first branch, $\beta_1 = \frac{72 y_0^2}{F_0}$, lets $\beta$ have a single zero, which as we saw after (\ref{eq:single0}) corresponds to a regular point. The second branch, $\beta_1=0$, makes $\beta$ have a double zero, which as we saw after (\ref{eq:double0}) corresponds to a D6. In this section we will use the first branch, leaving the second for section \ref{sub:gen}. 

Continuing to solve (\ref{eq:ode}) perturbatively after having set $\beta_1 = \frac{72 y_0^2}{F_0}$, we find a nice surprise: the perturbative expansion stops after three iterations. This leads to a very simple solution to (\ref{eq:ode}):
\begin{equation}\label{eq:nice}
	\beta= \frac8{F_0}(y-y_0)(y+2 y_0)^2\ .
\end{equation}
This has the desired single zero at $y=y_0$, and it also has a double zero at $y=-2y_0$, signaling that $M_3$ has a D6 stack there. These are the qualitative features one expects from the solution in \cite[Sec.~5.2]{afrt}; in that paper, that solution was argued to exist (along with many others, which we shall discuss in due course) on numerical grounds --- see in particular Fig.~3 in that paper. It would also be possible to find (\ref{eq:nice}) by finding the general solution, and imposing the presence of a simple zero; we will see this in section \ref{sub:gen}.

(\ref{eq:nice}) looks superficially very similar to (\ref{eq:betaMN}). Taking $c_1=-y_0^2/4$, we see that (\ref{eq:betaMN}) has \emph{two} double zeros, at $y=\pm y_0$, corresponding to two D6 stacks. This is indeed correct for that massless solution: the two D6 stacks are generated by the reduction from eleven dimensions, in a similar way as in \cite[Sec.~5.1]{afrt}. Notice also that the massless limit of (\ref{eq:nice}), on the other hand, does not exist, since $F_0$ appears there in the denominator.

Now that we have obtained one solution of (\ref{eq:nice}), we can pause to explore what the resulting AdS$_5$ solution looks like; moreover, using the map (\ref{eq:map57}), we can also produce an AdS$_7$ solution which will indeed be the one found numerically in \cite[Sec.~5.2]{afrt}. 

The conditions (\ref{eq:ineq}) give us two possibilities: 
\begin{equation}\label{eq:ineqF0}
	\{y_0 <0, \ F_0 >0, \ y \in [y_0,-2y_0] \} \ \ {\rm or} \ \ \{ y_0 >0, \ F_0 <0, \ y \in [-2y_0,y_0]\}\ .
\end{equation} 
We will assume the first possibility. One can then write the metric and fields most conveniently in terms of 
\begin{equation}
	\tilde y \equiv \frac{y}{y_0}\ ,
\end{equation}
which then has to belong to $[-2,1]$. We have
\begin{equation}\label{eq:metnice}
	ds^2_{M_5} = e^{2A} ds^2_{\Sigma_g} + \sqrt{-\frac{y_0}{8F_0}}\left(\frac{d\tilde y^2}{(1-\tilde y)\sqrt{\tilde y +2}} +\frac49 \frac{(1-\tilde y)(\tilde y+2)^{3/2}}{2-\tilde y} ds^2_{S^2}\right)\ ,
\end{equation} 
and
\begin{equation}\label{eq:Aphinice}
	e^{4A} = -2\frac{y_0}{F_0}(2+\tilde y) \ ,\qquad	e^{2 \phi}= \sqrt{-\frac1{2y_0 F_0^3}} \frac{(\tilde y +2)^{3/2}}{2-\tilde y}\ .
\end{equation} 
We also need to implement flux quantization, which in this case is the statement that the D6 stack at the $\tilde y = -2$ point has an integer number $n_2$ of D6-branes. This constraint was discussed right below (\ref{eq:b0}). From (\ref{eq:q5}) and (\ref{eq:nice}) we find $q_5=\frac13 \sqrt{2F_0 (y-y_0)}$, which implies 
\begin{equation}\label{eq:y0f0}
	y_0 = -\frac38 \frac{n_2^2}{F_0}\ .
\end{equation}
We did not replace this constraint in (\ref{eq:metnice}), as we did in (\ref{eq:nice5}), because later we will glue pieces of it together with other metrics and with itself, and in that context the parameter $y_0$ will be fixed by flux quantization a bit differently. 

The AdS$_7$ solutions can now be found easily by applying the map (\ref{eq:map57}), and in particular its action on the metric, (\ref{eq:57metric}). The internal metric on $M_3$ is
\begin{equation}\label{eq:met7nice}
	ds^2_{M_3} = \sqrt{-\frac{y_0}{6F_0}}\left(\frac{d\tilde y^2}{(1-\tilde y)\sqrt{\tilde y +2}} +\frac43 \frac{(1-\tilde y)(\tilde y+2)^{3/2}}{8 - 4 \tilde y - \tilde y^2} ds^2_{S^2}\right)\ ,
\end{equation}
and
\begin{equation}\label{eq:Aphi7nice}
	e^{4A}= -\left(\frac43\right)^3 2\frac{y_0}{F_0}(\tilde y + 2) \ ,\qquad
	e^{2 \phi}= \sqrt{-\frac 6{y_0 F_0^3}}\frac{(\tilde y +2)^{3/2}}{8-4\tilde y - \tilde y^2}\ .
\end{equation}
(\ref{eq:met7nice}) and (\ref{eq:Aphi7nice}) give analytically the solution found numerically in \cite[Sec.~5.2]{afrt}. The flux $F_2$ can be read off from the expression $F_2=q(\frac {x_7}4 F_0 e^{A+\phi}-1){\rm vol}_{S^2}$ in \cite[Eq.(4.42)]{afrt}:
\begin{equation}
	F_2=\frac{k}{\sqrt{3}}\frac{(1-\tilde y)^{3/2}(\tilde y +4)}{8-4\tilde y -\tilde y^2}{\rm vol}_{S^2}\ .
\end{equation}

For both the AdS$_5$ and AdS$_7$ solutions, from (\ref{eq:y0f0}) we can see that, making $n_2$ large, curvature and string coupling become as small as one wishes. This guarantees that the supergravity approximation is applicable. Similar limits can be taken for the solutions that we will present later. (This was shown in general in \cite[Sec.~4.1]{gaiotto-t-6d}.)


\subsection{General massive solution} 
\label{sub:gen}

Let us now go back to (\ref{eq:branch}) and see what happens if we use the branch $\beta_1=0$. This means that $\beta$ has a double zero, which corresponds to presence of a D6 stack at $y = -2y_0$. 

The perturbative expansion for (\ref{eq:ode}) now does not truncate anymore. It is possible to go to higher order, guess an expression for the $k$-th term $\beta_k$ in the Taylor expansion $\beta= \sum_{k=1}^\infty \beta_k (y-y_0)^k$, and resum this guess. (This is in fact the way we originally proceeded.) At this point it is of course much easier to use the trick explained below (\ref{eq:qtrick}), and find the general solution directly. Assuming $y_0>0$, it reads
\begin{equation}\label{eq:gen}
	\beta = \frac{y_0^3}{b_2^3 F_0}\left(\sqrt{\hat y} - 6\right)^2\left(\hat y + 6 \sqrt{\hat y}+6b_2 - 72 \right)^2 \ , 
\end{equation}
where
\begin{equation}
	\hat y \equiv 2 b_2 \left(\frac y{y_0} -1\right)+36\ ,\qquad b_2 \equiv \frac{F_0}{y_0} \beta_2 \ . 
\end{equation}
An alternative expression for (\ref{eq:gen}) is
\begin{equation}
	\sqrt{\beta} = \sqrt{\frac 8{F_0}} \sqrt{y- \tilde y_0}(y+ 2 \tilde y_0) -36 \sqrt{\frac{y_0^3}{b_2^3 F_0}} (b_2-12) \ ,
\end{equation}
where $\tilde y_0 = \left(1-\frac{18}{b_2}\right) y_0$. Notice the similarity with (\ref{eq:nice}).

This solution now depends on the two parameters $y_0$ and $b_2$, rather than just one as (\ref{eq:nice}), and we expect it to be the most general solution to (\ref{eq:ode}). To see whether this is true, let us analyze its features and compare them to what we expect from the qualitative study in \cite[Sec.~5.2]{afrt}; we will do so using the first expression (\ref{eq:gen}).

(\ref{eq:gen}) has zeros at $\hat y=36$ (which corresponds to $y=y_0$) and for $b_2 <12$ also at $\hat y = (-3 + \sqrt{81 - 6 b_2})^2$. 
Also, at $\hat y=0$ it has a point where it behaves as $\beta\sim \beta_0 + \sqrt{\hat y} + O(\hat y)$, which up to translation is the same as in (\ref{eq:sqrt}), which corresponds to an O6 point. Taking also into account the constraints in (\ref{eq:ineq}), we find two possibilities, and one special case between them. 
\begin{itemize}
	\item If $b_2<12$, the solution is defined in the interval $\hat y\in [(-3 + \sqrt{81 - 6 b_2})^2,36]$; there are two double zeros at both extrema. This represents a solution with two D6 stacks at both ends, but where the numbers of D6s are not the same on the two sides (unlike for (\ref{eq:betaMN})). Under the map (\ref{eq:map57}) to AdS$_7$, it becomes a solution that was briefly mentioned at the end of \cite[Sec.~5.2]{afrt}; in terms of the graph in Fig.~3(b) in that paper, its path would come from below and miss the green dot on the top side from the left, so as to end up in a D6 asymptotics on the top side as well.
	\item If $b_2>12$, the solution is defined for $\hat y\in [0,36]$; there is a double zero at $\hat y=36$, and an O6 singularity (see (\ref{eq:sqrt})) at $\hat y=0$. This represents a solution with one D6 stack at one end, and one O6 at the other extremum. Under the map to AdS$_7$, it becomes another solution that was briefly mentioned in \cite[Sec.~5.2]{afrt}; in terms of the graph in Fig.~3(b) in that paper, its path would come from below and miss the green dot on the top side from the right, so as to end up in an O6 asymptotics on the top side.
	\item In the limiting case, $b_2 = 12 $, the solution is again defined for $\hat y\in [0,36]$; under the map to AdS$_7$ we expect to find the case where (again referring to \cite[Fig.~3(b)]{afrt}) we hit the green dot at the top, which should correspond to having a regular point. Indeed in this case (\ref{eq:gen}) reduces to 
	\begin{equation}
		\beta = \frac{y_0^3}{1728 F_0} \hat y (\hat y -36)^2\ ,
	\end{equation}
	which has a double zero in $\hat y=36$ and a single zero in $\hat y=0$; it is essentially (\ref{eq:nice}). It would have been possible to obtain (\ref{eq:nice}) this way, but we chose to highlight it in a subsection by itself because of its simplicity.
\end{itemize}

So the solution (\ref{eq:gen}) has the features we expected from the qualitative analysis in \cite[Sec.~5.2]{afrt}. 

We record also here some data of the corresponding solutions. For the AdS$_5$ solution, the metric, warping and dilaton read
\begin{equation}
\begin{split}
	& 
	ds^2_{M_5} = e^{2A} ds^2_{\Sigma_g} + \frac{y_0^{5/4}  d\hat y^2}{4(b_2^5 F_0^3 \hat y^3 \beta)^{1/4}} + \frac{(b_2^7 F_0 \hat y)^{1/4}}{18 y_0^{7/4}}\frac{\beta^{3/4} ds^2_{S^2}}{2(b_2 -18)^2+18(b_2-12)\sqrt {\hat y} - (b_2-18)\hat y}\ ,  \\	
	&e^{8A}= \frac{b_2 \beta}{F_0 y_0 \hat y} \ ,\qquad
	e^{8 \phi} = \frac{b_2^{11} \beta^3}{16 F_0^3 y_0^{11} \hat y^3\left(2(b_2 -18)^2+18(b_2-12)\sqrt {\hat y} - (b_2-18)\hat y\right)^4}\ .
\end{split}
\end{equation}
The AdS$_7$ solution reads
\begin{equation}
\begin{split}
	& 
	ds^2_{M_3} = \frac{y_0^{5/4}  d\hat y^2}{4(b_2^5 F_0^3 \hat y^3 \beta)^{1/4}} + \frac{(b_2^7 F_0 \hat y)^{1/4}}{3 y_0^{7/4}}\frac{ \beta^{3/4} ds^2_{S^2}}{12(b_2 -18)^2+144(b_2-12)\sqrt {\hat y} - 12(b_2-18)\hat y-\hat y^2}\ ,  \\	
	&e^{8A}= \frac{2^{12}b_2 \beta}{3^6 F_0 y_0 \hat y} \ ,\qquad
	e^{8 \phi} = \frac{144 b_2^{11} \beta^3}{F_0^3 y_0^{11} \hat y^3\left(-12(b_2 -18)^2-144(b_2-12)\sqrt {\hat y} +12(b_2-18)\hat y + \hat y^2\right)^4}\ .
\end{split}	
\end{equation}

Finally, flux quantization can be taken into account by using (\ref{eq:b0}), (\ref{eq:b2O6}) and the expansion of $\beta$ around its zeros (or around its zero and its square root point, for the O6--D6 case). We obtain two equations, which discretize the two parameters $b_2$ and $y_0$. The expressions are not particularly inspiring (especially in the D6--D6 case) and we will not give them here.


\subsection{Some solutions with D8's} 
\label{sub:d8sol}

We will now show two simple examples of solutions with D8-branes. These will be the ones studied numerically in \cite[Sec.~5.3]{afrt}; here we will give their analytic expressions. We will simply have to piece together solutions we have already studied; all we will have to work out is the position of the D8's.

The first example is a solution with only one D8 stack. This can be obtained by gluing two metrics of the type (\ref{eq:metnice}). We will assume
\begin{equation}
	y_0 <0 \ ,\qquad F_0>0\ ; \qquad y_0'>0 \ ,\qquad F_0'<0\ .
\end{equation}
Following the logic in \cite[Sec.~5.3]{afrt}, the flux quantization conditions can be satisfied by taking for example the two-form flux integer after the D8 stack to vanish, $n_2'=0$, $n_2 = \mu (n_0'-n_0)$, $\mu\in \zz$, and  
\begin{equation}\label{eq:d81q}
	n_0' = n_0 \left(1-\frac N\mu\right)\ ,
\end{equation}
where $N= \frac1{4\pi^2}\int H$ is the NSNS flux integer. (Recall that $F_0= \frac{n_0}{2\pi}$, and similarly for $F_0'$.) As usual the metric can be written as $ds^2_{M_5}= e^{2A}ds^2_{\Sigma_g}+ ds^2_{M_3}$, and putting together two copies of (\ref{eq:metnice}) we can write\footnote{The sign differences between the expression before and after the D8 have to do with the simplification of factors involving $\sqrt{F_0^2}=|F_0|$ from applying (\ref{eq:Aphisol}) to (\ref{eq:nice}).}
\begin{equation}\label{eq:1d8}
	ds^2_{M_3}= \left\{ \begin{aligned}
\displaystyle 
    \frac1{\sqrt{8F_0}}\left(\frac{dy^2}{(y-y_0)\sqrt{-2y_0-y}}+ \frac49 \frac{(y-y_0)(-2y_0-y)^{3/2}}{-y_0 (y-2y_0)} ds^2_{S^2}\right)& , \quad  y_0<y<y_{{\rm D8}} \, ;  \\[7pt]
\displaystyle 
    \frac1{\sqrt{-8F_0'}}\left(\frac{dy^2}{(y_0'-y)\sqrt{2y_0'+y}}+ \frac49 \frac{(y_0'-y)(2y_0'+y)^{3/2}}{y_0' (2y_0'-y)} ds^2_{S^2}\right)& , \quad y_{\rm D8}<y<y_0'\ .
	\end{aligned}\right.
\end{equation}
We reverted to using $y$ rather than $\hat y$, so as to be able to use the same coordinate before and after the D8 stack. Imposing that $A$ and $\phi$ (or, equivalently, that $\beta$ and $\beta'$) be continuous across the D8 stack, we get  
\begin{equation}\label{eq:y00}
	y_0 = \frac12\frac{2 F_0 - F_0'}{F_0 + F_0'}y_{\rm D8} \ ,\qquad y_0' = \frac12\frac{2 F_0' - F_0}{F_0 + F_0'}y_{\rm D8}\ .
\end{equation}
We also have to impose (\ref{eq:qD8}), which fixes 
\begin{equation}
	y_{\rm D8}= y_0 +\frac{9 (F_0')^2 n_2^2}{8 F_0(F_0-F_0')^2}\ , 
\end{equation} 
which together with (\ref{eq:y00}) and (\ref{eq:d81q}) gives
\begin{equation}
	y_0 = -\frac32 F_0 \pi^2 (N^2-\mu^2) \ ,\qquad y_0'=\frac32 F_0 \pi^2 (N-\mu)(2 N- \mu)
	\ ,\qquad y_{\rm D8}= 3 F_0 \pi^2 (N-2\mu)(N-\mu)\ .
\end{equation}

One can also obtain the corresponding AdS$_7$ solution. This can be done using the map (\ref{eq:75metric}) on (\ref{eq:1d8}). Alternatively, we can just write one copy of (\ref{eq:met7nice}) for $y_0<y<y_{{\rm D8}}$, and a second copy of (\ref{eq:met7nice}), formally obtained by $y\to -y$, $y_0\to -y_0'$, $F_0 \to -F_0'$.
This provides the analytic expression of the solution in \cite[Fig.4]{afrt}. 

We can also consider a configuration with two D8 stacks. We will take it to by symmetric, in the sense that the flux integers before the first D8 stack will be $(n_0,0)$, between the two stacks $(0,n_2=-k<0)$, and after the second stack $(-n_0,0)$. This corresponds to \cite[Fig.~5]{afrt}. Again we will assume $y_0<0$; the positions of the two D8 stacks will be $y_{\rm D8}<0$ and $y_{\rm D8'}=-y_{\rm D8}>0$. We will give only the AdS$_7$ internal metric: 
\begin{equation}\label{eq:2D8}
	\hspace{-.5cm}ds^2_{M_3}= \left\{ \begin{aligned}
\displaystyle
		&\frac1{\sqrt{6F_0}}\left(\frac{dy^2}{(y-y_0)\sqrt{-2y_0-y}}+ \frac43 \frac{(y-y_0)(-2y_0-y)^{3/2}}{8y_0^2-4y y_0 -y^2} ds^2_{S^2}\right)
		 , &y_0<y<y_{\rm D8} \, ; \\[10pt]
\displaystyle
		&\frac{24^4 R^6 dy^2 +(9^2 R^6-32^2 y^2)^2 ds^2_{S^2}}{3 \cdot 6^5(9^2 R^6-32^2 y^2)^{1/2}}
		 \, , &y_{\rm D8}<y<-y_{\rm D8} \, ; \\[7pt]
\displaystyle
		&\frac1{\sqrt{6F_0}}\left(\frac{dy^2}{(-y_0-y)\sqrt{-2y_0+y}}+ \frac43 \frac{(-y_0-y)(-2y_0+y)^{3/2}}{8y_0^2+4y y_0 -y^2} ds^2_{S^2}\right)
		  &-y_{\rm D8}<y<-y_0 \, ; \\
	\end{aligned}\right.
\end{equation}
the metric in the middle region is the known massless metric in \cite[Eq.(5.4)]{afrt}, with the change of coordinate (\ref{eq:cosalpha}).

We now have three unknowns: $R$, $y_0$, $y_{\rm D8}$. Continuity of $\beta$ and $\beta'$ this time only imposes one condition; we then have (\ref{eq:qD8}) and the condition (\ref{eq:Hfluxq}). 
We get
\begin{equation}
\begin{split}
	y_0 = -\frac94 k &\pi (N-\mu) \ ,\qquad
	y_{\rm D8} = -\frac94 k \pi (N-2 \mu)\ ,\\
	&R^6 = \frac{64}3 k^2\pi^2 (3 N^2 -4\mu^2) \ ,\\
\end{split}	
\end{equation}
where in this case $\mu = \frac k{n_0}$. Notice that the in this case the bound in \cite[Eq.(4.10)]{gaiotto-t-6d} (which can also be found by (\ref{eq:Hfluxq})) implies $N>2\mu$.

It would now be possible to produce solutions with a larger number of D8's. It is in fact possible to introduce an arbitrary number of them, although there are certain constraints on their numbers and their D6 charges \cite[Sec.~4]{gaiotto-t-6d}. The most general solution can be labeled by the choice of two Young diagrams; there is also a one-to-one correspondence with the brane configurations in \cite{hanany-zaffaroni-6d,brunner-karch}. One can in fact think of the AdS$_7$ solutions as a particular near-horizon limit of the brane configurations. For more details, see \cite{gaiotto-t-6d}. For these more general solutions, we expect to have to glue together not only pieces of the solution in subsection \ref{sub:nice}  and of the massless solution, but also pieces of the more complicated solution in \ref{sub:gen}.


\subsection{Field theory interpretation} 
\label{sub:sum}

In this section we have found infinitely many new AdS$_5$ solutions in massive IIA, and we have established that they are in one-to-one correspondence with the AdS$_7$ solutions of \cite{afrt,gaiotto-t-6d}. 

It is easy to guess the field theory interpretation of this correspondence. 
Recall first the Maldacena--N\'u\~nez ${\cal N}=2$ solutions \cite{maldacena-nunez}. The original AdS$_7\times S^4$ solution of M-theory has an SO(5) R-symmetry; when one compactifies on a Riemann surface $\Sigma_g$, one ``mixes'' the SO(2) of local transformations on $\Sigma_g$ with an SO(2) $\subset$ SO(5) subgroup; the commutant SO(2)$\times$SO(3)$\cong$U(2) remains as the R-symmetry of the resulting ${\cal N}=2$ CFT$_4$. This is reflected in the form of the metric of the $S^4$, that gets distorted (except for the directions protected by the R-symmetry).

In similar ${\cal N}=1$ solutions \cite{maldacena-nunez,bah-beem-bobev-wecht}, the SO(2) is embedded in SO(5) in a more intricate way, so that its commutant is a U(1), which is indeed the R-symmetry of an ${\cal N}=1$ theory. 

For us, the CFT$_6$ has only $(1,0)$ supersymmetry, and thus its R-symmetry is already only SU(2). The twisting is very similar to the usual one in \cite{maldacena-nunez}: it is signaled by the fact that the $\psi$ coordinate is fibered over the Riemann surface $\Sigma_g$. 

When we mix this with the SO(2) of local transformations on $\Sigma_g$, the commutant is only a U(1). So in principle there is no symmetry protecting the shape of the internal $S^2$ in the AdS$_7$ solutions; indeed the metric (\ref{eq:metsol}) does not have SO(3) isometry, because the $\psi$ direction is fibered over $\Sigma_g$. What is a bit surprising is that the breaking is not more severe: (\ref{eq:S2}) might have become considerably more complicated, with $\sin\theta$ for example being replaced by a different function. Likewise, in the fluxes, one can see that there is no SO(3) symmetry: the $\cos\theta$ in front of ${\rm vol}_{\Sigma_g}$, for example, breaks it. Still, there are various nice ${\rm vol}_{S^2}$ terms which were not guaranteed to appear.

In any case, we interpret our solutions as the twisted compactification of the CFT$_6$ dual to the AdS$_7$ solutions in \cite{afrt,gaiotto-t-6d}. Recently, there has been a lot of progress in understanding such compactifications for the $(2,0)$ theories \cite{gaiotto,gaiotto-maldacena,bah-beem-bobev-wecht}, and it would be very interesting to extend those results to our AdS$_5$ solutions. Here, we will limit ourselves to pointing out a couple of preliminary results about the number of degrees of freedom. 

A common way of estimating the number of degrees of freedom using holography in any dimension is to introduce a cut-off in AdS, and estimate the Bekenstein--Hawking entropy (see for example \cite[Sec.~3.1.3]{magoo}). This leads to $\frac{R_{\rm AdS_7}^5}{G_{{\rm N},7}}$ in AdS$_7$, and to  $\frac{R_{\rm AdS_5}^3}{G_{{\rm N},5}}$ in AdS$_5$, where $G_{{\rm N},d}$ is Newton's constant in $d$ dimensions. The latter can be computed as $\frac1{g_s^2}{\rm vol}_{10-d}$. In a warped compactification with non-constant dilaton, both $R_{\rm AdS}$ and $g_s$ are non-constant, and should be integrated over the internal space. In our case, for AdS$_7$ this leads to 
\begin{equation}
	{\cal F}_{0,6} \equiv \int_{M_3} e^{5A_7-2 \phi_7}{\rm vol_3}
\end{equation}
and for AdS$_5$ to ${\cal F}_{0,4} \equiv \int_{M_5} e^{3A_5-2 \phi_5}{\rm vol_5}$. These can be thought of as the coefficient in the thermal partition function, ${\cal F}= {\cal F}_{0,d} V T^d$, where $V$ is the volume of space and $T$ is temperature. These computations however are basically the same for the coefficients in the Weyl anomaly, at least at leading order (i.e.~in the supergravity approximation). 

As a consequence of our map (\ref{eq:map57}), ${\cal F}_{0,6}$ and ${\cal F}_{0,4}$ are related. Taking into account the transformation of the volume form according to (\ref{eq:57metric}), we find
\begin{equation}\label{eq:FF}
	{\cal F}_{0,4}= \left(\frac34\right)^4 {\cal F}_{0,6} {\rm Vol}(\Sigma_g) \ .
\end{equation}
The volume of $\Sigma_g$ can be easily computed using Gauss--Bonnet and the fact that its scalar curvature equals $-6$: we get
\begin{equation}\label{eq:VolSigma}
	{\rm Vol}(\Sigma_g)= \frac43\pi (g-1)\ .
\end{equation} 
So the ratio of degrees of fredom in four and six dimensions is universal, in that it depends only on $g$ and not on the precise $(1,0)$ theory we are considering in our class. This is reminiscent of what happens for compactifications of the $(2,0)$ theory; see e.g.~\cite[Eq.(2.8)]{gaiotto-maldacena}, or \cite[Eq.~(2.22)]{bah-beem-bobev-wecht}.

\begin{figure}[ht]
\centering	
	\subfigure[\label{fig:ns5d6}]{\includegraphics[scale=.6]{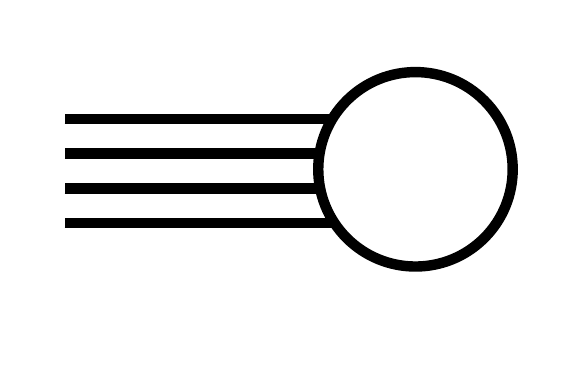}}
	\hspace{.4cm}
	\subfigure[\label{fig:ns5d6d8}]{\includegraphics[scale=.6]{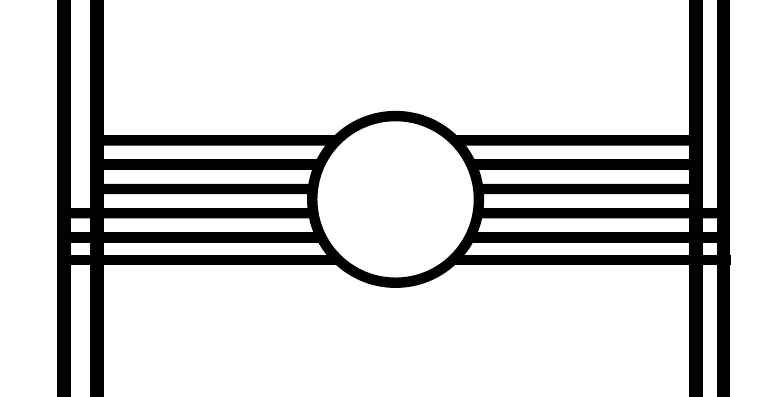}}
	\caption{Brane configurations for two sample theories. The circles represent stacks of $N$ NS5-branes; the horizontal lines represent D6-branes; the vertical lines represent D8-branes. In the second case, on each side we have $n_0=2$ D8-branes; $|\mu|=3$ D6-branes end on each, for a total of $k= n_0 |\mu|=6$.}
	\label{fig:ns5}
\end{figure}

We have not computed ${\cal F}_{0,6}$ in full generality for the $(1,0)$ theories. This would now be possible in principle, since the analytic expressions are now known. One first example is the solution in section \ref{sub:nice}. The corresponding brane configuration according to the identification in \cite{gaiotto-t-6d} consists in $k$ D6's ending on $N=\frac k{n_0}$ NS5-branes; see figure \ref{fig:ns5d6}. We get 
\begin{equation}
	{\cal F}_{0,6}= \frac{512}{45} k^2\pi^4 N^3\ ,
\end{equation} 
which reassuringly goes like $N^3$. (By way of comparison, for the massless case one gets ${\cal F}_{0,6}=\frac{128}3 k^2 \pi^4 N^3$.)

We also computed ${\cal F}_{0,6}$ for the solution (\ref{eq:2D8}), which has two D8's and a massless region between them. The corresponding brane configuration would be $N$ NS5-branes in the middle with $k=\mu n_0$ D6's sticking out of them, ending on $n_0$ D8-branes both on the left and on the right; see figure \ref{fig:ns5d6d8}.  This case was considered in \cite[Sec.~5]{gaiotto-t-6d}, where approximate expressions for ${\cal F}_{0,6}$ were computed, using perturbation theory around the massless limit. Using (\ref{eq:2D8}) we can now obtain the exact result: 
\begin{equation}
	{\cal F}_{0,6}= \frac{128}3 k^2\pi^4 \left(N^3 - 4 N\mu^2 + \frac{16}5 \mu^3\right)\ .
\end{equation}
This agrees with \cite[Sec.~5]{gaiotto-t-6d}, but is now exact. Recall that $\mu=\frac k{n_0}$; since this number can be large, the second and third term are also large, and are not competing with stringy corrections. Using (\ref{eq:FF}), and comparing with the $(2,0)$ theory to fix the proportionality factors, we get that for the CFT$_4$ theory $a=c= \frac13(g-1)\left(N^3 - 4 N\mu^2 + \frac{16}5 \mu^3\right)$. Stringy corrections will modify this result with terms linear in $N$ and probably in $\mu$.




\section{Conclusions} 
\label{sec:conc}

We have classified supersymmetric AdS$_5 \times M_5$ solutions of massive type IIA supergravity, and we have found a large class of new analytic solutions. 

The general classification, obtained in section \ref{generalanalysis}, is summarized in section \ref{sub:summary}. We reduced the supersymmetry equations to six PDE's. A solution to this system completely determines the bosonic fields --- metric, dilaton, and fluxes. The geometry of $M_5$ is given by a fibration of a three-dimensional manifold $M_3$ over a two-dimensional space ${\cal C}$.

We found an Ansatz that makes the PDE system solvable. As described in section \ref{compAnsatz}, it consists in relating the metric on ${\cal C}$ to the warping function $A$. We recover in this way several known massless solutions: the BBBW \cite{bah-beem-bobev-wecht0,bah-beem-bobev-wecht}, Maldacena--N\'u\~nez \cite{maldacena-nunez}, and INST \cite{itsios-nunez-sfetsos-thompson} solutions. More interestingly, we find new analytic solutions. 

This new class, analyzed in section \ref{sec:sol}, consists of infinitely many new solutions, which preserve eight supercharges in five dimensions and are in one-to-one correspondence with the AdS$_7 \times S^3$ type IIA backgrounds classified in \cite{afrt}. We have explicitly described the map between the former and the latter. The geometry of the fibre $M_3$ inside $M_5$ is a certain modification of the ``distorted'' $S^3$ of the AdS$_7$ compactifications, whereas the base ${\cal C}$ is a Riemann surface with constant negative curvature and genus $g>1$. An $S^2$ inside $M_3$ is twisted over ${\cal C}$, breaking the SU(2)-isometry of $M_3 \cong S^3$ to U(1); this bears out the field-theoretic expectation of having a U(1) R-symmetry for the dual four-dimensional $\mathcal{N}=1$ SCFT.

Importantly, we have been able to find analytic expressions for all these AdS$_5$ solutions. Then, by means of the aforementioned one-to-one correspondence, we have obtained analytic versions of all the AdS$_7$ solutions in \cite{afrt} (which were previously known only numerically). Thanks to the analytic expressions for the fields on the gravity side, we have computed explicitly the free energy for some examples of four-dimensional $\mathcal{N}=1$, and six-dimensional $\mathcal{N}=(1,0)$ SCFT's at large $N$ (using the AdS/CFT dictionary). It would be very interesting to find a field theory description of these theories, perhaps along the lines of \cite{gaiotto}.



\section*{Acknowledgments}
We would like to thank C.~Bachas, I.~Bah, I.~Bena, N.~Bobev, A.~Zaffaroni for interesting discussions. We are especially indebted with A.~Rota, who made several interesting comments towards the end of this paper, when the similarities with \cite{rota-t} became apparent.  F.A.~is grateful to the Graduiertenkolleg GRK 1463 ``Analysis, Geometry and String Theory'' for support. The work of M.F.~was partially supported by the ERC Advanced Grant ``SyDuGraM'', by IISN-Belgium (convention 4.4514.08) and by the ``Communaut\'e Fran\c{c}aise de Belgique" through the ARC program. M.F.~is a Research Fellow of the Belgian FNRS-FRS. A.P.~and A.T.~are supported in part by INFN and by the European Research Council under the European Union's Seventh Framework Program (FP/2007-2013) -- ERC Grant Agreement n. 307286 (XD-STRING). The research of A.T.~is also supported by the MIUR-FIRB grant RBFR10QS5J ``String Theory and Fundamental Interactions''.

\appendix

\section{Supersymmetry variations and the Killing vector}
\label{susyvar}

Setting to zero the type IIA supersymmetry variations (of gravitinos and dilatinos) yields the following set of equations\footnote{The first two equations follow from setting the gravitino variation $\delta\psi_M$ to zero, while the last two equations follow from $\Gamma^M \delta\psi_M - \delta\lambda = 0$ where $\lambda$ is the dilatino.}
\begin{subequations}\label{10dSUSYvariations}
\begin{align}
0 &= \left(\nabla_M + \frac{1}{4} H_M\right)\epsilon_1 + \frac{e^{\phi}}{16} \lambda(F) \Gamma_M \epsilon_2 \ , \\
0 &= \left(\nabla_M - \frac{1}{4} H_M\right)\epsilon_2 + \frac{e^{\phi}}{16} F \, \Gamma_M \epsilon_1 \ , \\
0 &= \left(\nabla - \partial \phi + \frac{1}{4} H \right) \epsilon_1 \ , \\
0 &= \left(\nabla - \partial \phi - \frac{1}{4} H \right) \epsilon_1 \ ,
\end{align} 
\end{subequations}
where suppressed indices are contracted with antisymmetric products of gamma matrices and $\epsilon_1$, $\epsilon_2$ are Spin$(1,9)$ Majorana--Weyl spinors of opposite chirality. 

We wish to obtain a  set of differential and algebraic equations for the Spin$(5)$ spinors $\eta_1$, $\eta_2$ and so we decompose the the generators of Cliff$(1,9)$ as
\begin{equation}
\Gamma_\mu = e^A \gamma^{(1,4)}_\mu \otimes 1 \otimes \sigma_3 \, \qquad
\Gamma_i = 1 \otimes \gamma^{(5)}_m \otimes \sigma_1 \ , 
\end{equation}
where $\mu = 0, \dots, 4$, $m = 1, \dots, 5$ and $\sigma_1$ and $\sigma_3$ are the Pauli matrices; $\gamma^{(1,4)}_\mu$ generate Cliff$(1,4)$ and $\gamma^{(5)}_m$ Cliff$(5)$. Accordingly, the chirality matrix $\Gamma_{11}$ and the intertwiner $B_{10}$ relating $\Gamma_M$ and $\Gamma_M^*$,  are decomposed as
\begin{equation}
\Gamma_{11} = 1 \otimes 1 \otimes \sigma_2 \ , \qquad B_{10}  = B_{1,4} \otimes B_5 \otimes \sigma_1 \ .
\end{equation}
Furthermore, the supersymmetry parameters $\epsilon_1$, $\epsilon_2$ split as
\begin{subequations}
\begin{align}
\epsilon_1 &= (\zeta \otimes \eta_1 + \zeta^c \otimes \eta_1^c) \otimes \theta \ , \\
\epsilon_2 &= (\zeta \otimes \eta_2 + \zeta^c \otimes \eta_2^c) \otimes \theta^* \ ,
\end{align}
\end{subequations}
where $\eta^c_{1,2} = B_5 \eta_{1,2}^*$ and $\zeta^c = B_{1,4}\zeta^*$. $\zeta$ is a Spin$(1,4)$ spinor obeying the AdS$_5$ Killing spinor equation
\begin{equation}
\nabla_\mu \zeta = \frac{1}{2} \gamma_\mu \zeta \ ,
\end{equation}
while $\theta$ obeys $\sigma_2 \theta = \theta$ and $\sigma_1 \theta = \theta^*$. 

Applying the above decomposition, the equations \eqref{10dSUSYvariations} become 
\begin{subequations}\label{5dSUSYvariations}
\begin{align}
0 &= \left(\nabla_i + \frac{1}{4} H_i\right)\eta_1 
+ \frac{e^{\phi}}{16} \lambda(F) \gamma^{(5)}_m \eta_2 \ , \label{5dSUSYvariationsdiff1} \\
0 &= \left(\nabla_i - \frac{1}{4} H_i\right)\eta_2 
+ \frac{e^{\phi}}{16} F \, \gamma^{(5)}_m \eta_1 \ , \label{5dSUSYvariationsdiff2} \\
0 &= \left(\frac{i}{2} e^{-A} - \frac{1}{2} \partial A \right)\eta_1 
- \frac{e^\phi}{16} \lambda(F) \eta_2  \ , \label{5dSUSYvariationsalg1} \\
0 &= \left(\frac{i}{2} e^{-A} + \frac{1}{2} \partial A \right)\eta_2 
+ \frac{e^\phi}{16} F \eta_1 \ , \label{5dSUSYvariationsalg2} \\
0 &= \left(
\frac{5i}{2} e^{-A} - \nabla - \frac{5}{2} \partial A  + \partial \phi - \frac{1}{4} H 
\right)\eta_1 \ , \label{5dSUSYvariationsalg3}\\
0 &= \left(
\frac{5i}{2} e^{-A} + \nabla + \frac{5}{2} \partial A  - \partial \phi - \frac{1}{4} H 
\right)\eta_2 \ .\label{5dSUSYvariationsalg4}
\end{align}
\end{subequations}
Using equations \eqref{5dSUSYvariationsdiff1} and \eqref{5dSUSYvariationsdiff2} it is straightforward to show that $\xi \equiv  \frac{1}{2}(\eta_1^\dagger \gamma^m \eta_2 - \eta_2^\dagger \gamma^m \eta_2) \partial_m$ satisfies
\begin{equation}
\nabla_{(m} \, \xi_{n)} = 0 \ ,
\end{equation}
i.e.\ that $\xi$ is a Killing vector, while equations \eqref{5dSUSYvariationsalg1} and \eqref{5dSUSYvariationsalg2} yield $\mathcal{L}_\xi A = 0$. That $\mathcal{L}_\xi \phi = 0$ follows from the algebraic equations obtained from \eqref{5dSUSYvariationsalg3} and \eqref{5dSUSYvariationsalg4} afer eliminating $\nabla$, using \eqref{5dSUSYvariationsdiff1} and \eqref{5dSUSYvariationsdiff2}.\footnote{These conditions also follow directly from setting the dilatino variation $\delta\lambda$ to zero.}

\section{A simple Ansatz}
\label{simpleAnsatz}

We assume that $\phi$ and $A$ are functions of $y$ only and that $g_\mathcal{C}$ is independent of $x$ i.e.\ $f_1 = 0$.
From equation \eqref{fx} it follows that $\ap_2 = 0$. The metric becomes
\begin{equation}\label{eq:metAns1}
ds^2_{M_5} = ds^2_{\mathcal{C}} + \frac{1}{9} e^{2A} b^2 D\psi^2 + e^{-8A + 2\phi} dx^2 + \frac{ e^{-4A + 2\phi}}{b^2} dy^2 \ ,
\end{equation}
where now $b^2 = 1 - \ap_1^2$. Equation \eqref{fxfyinteg} is satisfied trivially while equations \eqref{dxrho0} and \eqref{dyrho0} yield $\partial_x \rho = \partial_y \rho = 0$ (in the present Ansatz $\rho_0 = \rho$). $A(y)$ and $\phi(y)$ are subject to the differential equations coming from the Bianchi identities of $F_0$ and $F_2$, and equation \eqref{dyell},
\begin{equation}\label{simpleAnsatzdyell}
\partial_y \ell + 2 f_2 \ell = 0 \ .
\end{equation}
$\ell$ is determined by \eqref{d2rho0} and \eqref{d2rho} to be $\ell = 6 e^{-2A} + 12 e^{-2A} y(\partial_y A - f_2)$.

We first look at the Bianchi identity of $F_2$; it yields:
\begin{equation}
F_0 (\partial_y A - f_2) = 0 \ ,
\end{equation}
so either $F_0 = 0$ or $f_2 = \partial_y A$. We consider the two cases $F_0 = 0$ and $F_0 \neq 0$ separately.

\subsection{\texorpdfstring{$F_0 = 0$}{F(0) = 0}}
\label{simpleAnsatzF0zero}
In this case, from the expression $\eqref{F0}$ for $F_0$ we conclude that $\partial_y A = 0$ i.e.\ $A$ is constant which without loss of generality we set to zero. $F_2$ is zero, as can be seen from its expression \eqref{F2}. We thus need to solve equation \eqref{simpleAnsatzdyell}. This yields the ODE
\begin{equation}
\partial_y f_2 + 2 f_2^2 = 0 \ ,
\end{equation}
which is solved by
\begin{equation}\label{simpleAnsatzfysol}
f_2 = \frac{1}{2}\frac{c}{cy - k} \ , \qquad c,\, k = \mathrm{const.} \ .
\end{equation}
Recalling the definition \eqref{fy} of $f_2$, equation \eqref{simpleAnsatzfysol} is in turn solved for
\begin{equation}
e^{2\phi} = \frac{k - cy}{2(c_1 - ky^2)} \ , \qquad c_1 = \mathrm{const.} \ .
\end{equation}
Equations \eqref{dudecomp} are then solved by
\begin{equation}\label{simpleAnsatzu}
u = e^{i\psi} \sqrt{2(k - c y)} \, \widehat u (x_1,x_2) \ .
\end{equation}
Substituting \eqref{simpleAnsatzfysol} and \eqref{simpleAnsatzu} in \eqref{d2rho} yields
\begin{equation}
d_2 \rho =  12 k \, \vol_\Sigma \ ,
\end{equation}
where $\Sigma$ is the surface spanned by $\widehat u$. Its Gaussian curvature is thus $12k$.

This solution was first discovered by Gauntlett, Martelli, Sparks and Waldram \cite{gauntlett-martelli-sparks-waldram-M} (see appendix \ref{GMSW}), and it is the T-dual of the AdS$_5 \times Y^{p,q}$ solution in type IIB supergravity.  

\subsection{\texorpdfstring{$F_0 \neq 0$}{F(0) != 0}}
In this case $f_2 = \partial_y A$; equations \eqref{dudecomp} are solved by
\begin{equation}
u = e^{i\psi} e^A \, \widehat u (x_1,x_2) \ .
\end{equation}
$\ell = 6 e^{-2A}$ obeys equation \eqref{simpleAnsatzdyell} automatically and 
\begin{equation}
d_2 \rho = 6 \vol_\Sigma \ .
\end{equation}

Substituting $f_2$ in \eqref{fx} gives 
\begin{equation}\label{eq:Ans1-1}
e^{4A} = - \frac{1}{12y} \partial_y \beta \ ,
\end{equation}
where $\beta(y) \equiv e^{10A-2\phi} b^2 $. The Bianchi identity of $F_0$ becomes then an ODE for $\beta$:
\begin{equation}\label{eq:Ans1-2}
e^{12A} F_0 = - \beta \, \partial_y e^{4A}   \ .
\end{equation}

The situation appears promising: we have reduced the problem to the ODE (\ref{eq:Ans1-2}). However, as we will now see, one cannot obtain physical compact solutions to this system. 

Let us introduce the coordinate $\tilde y $ by $d \tilde y = \frac{e^{-2A + \phi}}b dy$, so that the metric (\ref{eq:metAns1}) contains $d\tilde y^2$. (\ref{eq:Ans1-1}) now reads 
\begin{equation}\label{eq:Ans1-yt}
	F_0 = 16 e^{-\phi} b^2 \del_{\tilde y} A \ .
\end{equation}
In order to obtain a compact solution, we should have the factor in front of the $S^1$ in (\ref{eq:metAns1}), namely $e^A b$, go to zero for some $y=y_0$. For a regular point, this is impossible: since $A$ and $\phi$ should go to constant at $y_0$, we should have $b$ go to zero; but from (\ref{eq:Ans1-yt}) we see that this is in contradiction with $F_0\neq 0$. We might think of having a singularity corresponding to a brane, but since only an $S^1$ would shrink at $y=y_0$, such a brane would be codimension-2; there are no such objects in IIA supergravity.

\section{Recovered solutions}
\label{recoveredsolutions}

In this appendix we discuss a set of known, supersymmetric AdS$_5 \times M_5$ solutions of type IIA supergravity with zero Romans mass, which we recovered in our analysis. Two of them descend from AdS$_5$ solutions of M-theory, whose reduction to ten dimensions we present. We focus on the geometry of the solutions, as the fluxes are determined by it. We aim to adhere to the notation of the original papers; whenever there is overlap with notation used in the main body of the paper, we add a hat $\hat{} \ $.

There are more supersymmetric AdS$_5$ solutions in IIA \cite{aharony-berdichevsky-berkooz,sfetsos-thompson,sfetsos-thompson2,reidedwards-stefanski} that should be particular cases of our general classification of section \ref{generalanalysis}. These are outside the compactification Ansatz of section \ref{compAnsatz}.

\subsection{The Gauntlett--Martelli--Sparks--Waldram (GMSW) solution}
\label{GMSW}

The metric on $M_5$ reads 
\begin{align}
ds^2_{M_5} &= \frac{k-cy}{6m^2} ds^2_{C_k} + e^{-6\lambda} \sec^2\zeta + \frac{1}{9m^2}\cos^2\zeta D\psi^2 + e^{-6\lambda} dx_3^2 \ , 
\end{align}
where
\begin{equation}
e^{6\lambda} = \frac{2m^2(\hat a - k y^2)}{k - c y} \ , \qquad 
\cos^2\zeta = \frac{\hat a - 3ky^2 + 2cy^2}{\hat a - k y^2} \ . 
\end{equation}
The dilaton is given by $e^{-2\phi} = e^{6\lambda}$.

$\hat a$, $c$ are constants, $k = 0, \pm 1$ and $m^{-1}$ is the radius of AdS$_5$. $C_k$ is a Riemann surface of unit radius; it is a sphere $S^2$, a torus $T^2$ or a hyperbolic space $H^2$ for $k = 1, 0$ or $-1$ respectively. The GMSW solution is the reduction to ten dimensions of an AdS$_5 \times M_6$ solution of M-theory, where $M_6$ is a fibration of $S^2$ over $C_k \times T^2$ and the reduction is along an $S^1 \in T^2$.

The solution is the one recovered in subsection \ref{simpleAnsatzF0zero}. The constants $c$ and $k$ are identified with the corresponding of \ref{simpleAnsatzF0zero}, while $\hat a = c_1$. The coordinate $x_3$ is related to $x$ via $x_3 = - x$; a minus is introduced for matching the expressions of the fluxes. Finally, in \ref{simpleAnsatzF0zero} $m=1$.

\subsection{The Itsios--N\'{u}\~{n}ez--Sfetsos--Thompson (INST) solution}
\label{INST}

The INST solution \cite{itsios-nunez-sfetsos-thompson} was discovered by nonabelian T-dualizing the AdS$_5 \times T^{1,1}$ solution in type IIB supergravity. The metric on $M_5$ reads 
\begin{align}
ds^2_{M_5} &= \lambda_1^2 ds^2_{S^2} + \frac{\lambda_2^2 \lambda^2}{\Delta} x_1^2 D\psi^2 + \frac{1}{\Delta} \Bigl[ (x_1^2 + \lambda^2 \lambda_2^2 ) dx_1^2 + (x_2^2 + \lambda_2^4) dx_2^2 + 2x_1 x_2 dx_1 dx_2 \Bigr] \ , 
\end{align}
where
\begin{equation}
\Delta = \lambda_2^2 x_1^2 + \lambda^2 (x_2^2 + \lambda_2^4) \ , \qquad
\lambda_1^2 = \lambda_2^2 = \frac{1}{6} \ ,\qquad \lambda^2 = \frac{1}{9} \ ,
\end{equation}
and
\begin{equation}
ds^2_{S^2} = d\theta_1^2 + \sin^2\theta_1 \phi_1^2 \ , \qquad \rho = \cos\theta_1 d\phi_1 \ .
\end{equation}
The dilaton is given by $e^{-2\phi} = \Delta$.

The INST solution fits into the $c_1 = 0$ branch of the first case of subsection \ref{compAnsatzF00} for $c_3 = -12$ (achieved by setting the constant warp factor to zero) and $\epsilon = c_2 = \lambda\lambda^2_2$. $\Sigma_g$ is $S^2$ of radius $\frac{1}{\sqrt{6}}$. The coordinate transformation relating $x_1, x_2$ to $x, y$ is:
\begin{align}
x_1^2 = - 36 y^2 + 36 \epsilon x + 6 c_4 - 6 \epsilon^2 \ , \qquad x_2 = 6 y \ .
\end{align}

\subsection{The Maldacena--N\'{u}\~{n}ez solution}
\label{MN}

We write the metric of the $\mathcal{N} = 1$ Maldacena--N\'{u}\~{n}ez solution \cite{maldacena-nunez} in the form presented in \cite{gauntlett-martelli-sparks-waldram-M}:
\begin{equation}
e^{-2\lambda} ds^2_{11} = ds^2_{\mathrm{AdS}_5} + \frac{1}{3}ds^2_{H^2} + e^{-6\lambda} \sec^2\zeta dy^2 
+ \frac{1}{9m^2} \cos^2\zeta \left((d\psi + \tilde{P})^2 + ds^2_{S^2} \right)\ ,
\end{equation}
where
\begin{equation}
e^{6\lambda} = \hat a + y^2 \ , \qquad 
\cos^2\zeta = \frac{\hat a-3y^2}{\hat a + y^2} \ ,
\end{equation}
and $m^{-1}$ is the radius of AdS$_5$. 
The metrics on $H^2$ and $S^2$ are
\begin{equation}
ds^2_{H^2} = \frac{dX^2 + dY^2}{Y^2} \ , \qquad
ds^2_{S^2} = d\theta^2 + \sin^2\theta d\nu^2 \ ,
\end{equation}
while the connection of the fibration of $\psi$ is 
\begin{equation}
\tilde P = -\cos\theta d\nu - \frac{dX}{Y} \ .
\end{equation}

\subsubsection{Reduction to ten dimensions} 
\label{ssub:mn}

We reduce the Maldacena--N\'{u}\~{n}ez solution to ten dimensions, along $\nu$. In order to do so,
we rewrite the part of $ds^2_{M_6}$ involving $d\psi$ or $d\nu$ as
\begin{equation}
\frac{1}{9m^2} \cos^2\zeta \left[ (d\nu + A_1)^2 + \sin^2\theta D\psi^2 \right] \ ,
\end{equation}
where
\begin{equation}
A_1 = - \cos\theta D\psi \ , \qquad \rho = - \frac{dX}{Y} \ . 
\end{equation}
Reducing along $d\nu$ yields then
\begin{equation}\label{reducedMN}
ds^2_{10} = e^{2A}ds^2_{\mathrm{AdS}_5} + ds^2_{M_5}\ ,
\end{equation}
where
\begin{equation}\label{reducedMN_M5metric}
 e^{-2A} ds^2_{M_5} = \frac{1}{3}ds^2_{H^2} + e^{-6A+2\phi} \sec^2\zeta dy^2 + \frac{1}{9m^2} \cos^2\zeta \left( d\theta^2 + \sin^2\theta D\psi^2 \right)\ .
\end{equation}
Furthermore,
\begin{equation}
\phi = \frac{3}{4} \log \left( \frac{1}{9m^2} e^{2\lambda} \cos^2\zeta\right) \ , \qquad
A = \lambda + \frac{1}{3}\phi \ .
\end{equation}

The reduced Maldacena--N\'{u}\~{n}ez solution fits into the second case of section \ref{compAnsatzF00}, for $\epsilon = 0$ (achieved by by a $x \rightarrow x + \frac{\epsilon}{3}$ shift), $c_1 = - \frac{\hat a}{12}$ and $c_2 = 1$. $\Sigma_g$ is $H^2$ of radius $\frac{1}{\sqrt{3}}$. In our conventions $m=1$. The coordinate transformation relating $x$ to $y, \theta$ is:
\begin{equation}
x = -\frac{1}{9} (\hat a - 3 y^2) \cos\theta \ . 
\end{equation}


\subsubsection{\texorpdfstring{AdS$_7$ variables}{AdS(7) variables}} 
\label{ssub:mnads7}

For our discussion in the main text, it is useful to also include two parameters $R$ and $k$ which are usually set to one. If we use the slightly awkward-looking
\begin{equation}
	\beta = \frac4{k^2}\left(y^2- \frac{3^4}{2^{10}}R^6\right)^2
\end{equation}
the corresponding solution, using (\ref{eq:metsol}) and (\ref{eq:Aphisol}), is 
\begin{align}
	&ds^2_{M_5}=e^{2A}ds^2_{\Sigma_g} + \frac{1}{3^{3/2}k}\frac{64 dy^2}{\sqrt{9^2 R^6 - 32^2 y^2}}+ \frac{(9^2 R^6 - 32^2 y^2)^{3/2}}{16(3^5 R^6+ 32^2 y^2)}\ ,\\
	&e^{4A}= \frac{9^2 R^6 - 32^2 y^2}{3\cdot 2^8 k^2} \ ,\qquad e^{4 \phi} = \frac{(9^2 R^6 - 32^2 y^2)^3}{2\cdot 6^3 k^6 (3^5 R^6+ 32^2 y^2)^2}\ .
\end{align}
These again look messy, but upon using the map (\ref{eq:map57}) and defining an angle $\alpha$ via
\begin{equation}\label{eq:cosalpha}
	\cos\alpha \equiv \frac{32}{9 R^3} y 
\end{equation}
turn into the expressions for the metric, $A$ and $\phi$ of the massless AdS$_7$ solution, obtained by reducing AdS$_7\times S^4/\zz_k$ to IIA supergravity: see \cite[Sec.~5.1]{afrt}.

In the main text we will need an expression for the $B$ field of the AdS$_5$ solution. We give it directly in terms of $x_7$, which is related to (\ref{eq:x5}) via (\ref{eq:map57}):
\begin{equation}\label{eq:Bmn}
	B= \frac{R^3}{48k}x_7 \frac{(5-x_7^2)}{1+\frac13 x_7^2} {\rm vol}_{S^2}+\frac1{\sqrt{3}}\frac{x_7}{\sqrt{1-x_7^2}} \cos\theta {\rm vol}_{\Sigma_g}\ .
\end{equation}
This is similar to the one given for the AdS$_7$ solution in \cite[Eq.(5.8)]{afrt}.


\subsection{The Bah--Beem--Bobev--Wecht (BBBW) solution}
\label{BBBW}

The metric of the BBBW \cite{bah-beem-bobev-wecht} solution is
\begin{equation}
ds^2_{11} = e^{2\lambda} \left[ ds^2_{\mathrm{AdS}_5} +  e^{2\nu +2\hat A(x_1,x_1)} (dx_1^2+dx_2^2) \right] + e^{-4\lambda} ds^2_{M_4} \ , 
\end{equation}
where $ds^2_{\mathrm{AdS}_5}$ is the unit radius metric on AdS$_5$, and $\hat A(x_1,x_2)$ 
is the conformal factor of the constant curvature metric on the Riemann surface $\widehat\Sigma_g$ of genus $g$, obeying
\begin{equation}
(\partial^2_{x_1} + \partial^2_{x_2} )\hat A + \kappa e^{2\hat A} = 0 \ .
\end{equation}
The constant $\kappa$ is the Gaussian curvature of the Riemann surface which is set to $1$, $0$ or $-1$ for the sphere $S^2$, the torus $T^2$ or a hyperbolic surface respectively. $\nu$ is a real constant. The metric $ds^2_{M_4}$ is
\begin{align}
ds^2_{M_4} = \left( 1+ \frac{4 y^2}{q f} \right) dy^2 + \frac{q f}{k} \left(dq + \frac{12  y k}{q f} dy\right)^2 
+ \frac{\hat a_1^2}{4}~ \frac{f k}{q}(d\chi +V)^2 + \frac{q f}{9} (d\psi + \hat \rho)^2 \ .
\end{align}
The metric functions are
\begin{align}
e^{6\lambda} = q f +4y^2 \ , \qquad
f(y) \equiv 1+ 6 \frac{\hat a_2}{\hat a_1} y^2 \ , \qquad
k(q) \equiv \frac{\hat a_2}{\hat a_1} q^2 + q - \frac{1}{36} \ ,
\end{align}
while the one-forms which determine the fibration of the $\psi$ and $\chi$ directions are given by
\begin{align}
\hat \rho = (2-2g) V - \frac{1}{2}\left(\hat a_2 + \frac{\hat a_1}{2q}\right) (d\chi + V) \ , \qquad
dV =\frac{\kappa}{2-2g} e^{2\hat A} dx_1 \wedge dx_2 \ .
\end{align}
The constants $\hat a_1$, $\hat a_2$ are fixed as 
\begin{equation}
\hat a_1 \equiv \frac{2(2-2g)e^{2\nu}}{\kappa} \ , \qquad  
\hat a_2 \equiv 2(2-2g) \left(1 - \frac{6 e^{2\nu}}{\kappa}\right) \ . 
\end{equation}

\subsubsection{Reduction to ten dimensions}

We reduce the BBBW solution to ten dimensions, along $\chi$. In order to do so,
we rewrite the part of $ds^2_{M_4}$ involving $d\psi$ or $d\chi$ as
\begin{equation}
h_1^2 (d\chi + A_1)^2 + h_2^2 D\psi^2 \ ,
\end{equation}
where
\begin{subequations}
\begin{align}
h_1^2(y,q) &\equiv \frac{\hat a^2_1}{4} \frac{f k}{q} + \frac{1}{4} \frac{q f}{9} \left(\hat a_2 + \frac{\hat a_1}{2q}\right)^2 \ , \\
h_2^2(y,q) &\equiv \frac{q f}{9} -\frac{1}{4}  \left(\frac{q f}{9}\right)^2 \left(\hat a_2 + \frac{\hat a_1}{2q}\right)^2 \ ,
\end{align}
\end{subequations}
and
\begin{equation}
\rho = (2-2g) V \ , \qquad A_1 = V - \frac{qf}{9} \frac{1}{2} \left(\hat a_2 + \frac{\hat a_1}{2q}\right) h_1^{-1}  D\psi \ . 
\end{equation}
Reducing along $d\chi$ yields then
\begin{equation}\label{reducedBBBBW}
ds^2_{10} = e^{2A}\left[ds^2_{\mathrm{AdS}_5} + e^{2\nu +2\hat A} (dx_1^2+dx_2^2)\right] + ds^2_{M_3}\ ,
\end{equation}
where
\begin{equation}\label{reducedBBBW_M3metric}
e^{4A-2\phi} ds^2_{M_3} = \left( 1+ \frac{4 y^2}{q f} \right) dy^2 + \frac{q f}{k} \left(dq + \frac{12  y k}{q f } dy\right)^2 + h_2^2 D\psi^2 \ .
\end{equation}
Furthermore,
\begin{equation}
\phi = \frac{3}{2} (\log h_1 - 2\lambda) \ , \qquad
A = \frac{1}{2} \log h_1 \ .
\end{equation}
The reduced BBBW solution fits into the generic branch of the first case of subsection \ref{compAnsatzF00} for $c_1 = \frac{9 \hat a_1 + \hat a_2}{108}$, $c_2 = \frac{(9 \hat a_1 + \hat a_2)\hat a_2}{18 \hat a_1}$ and $c = \frac{\hat a_2}{3\hat a_1}$. The coordinate transformation relating $x$ to $y, q$ is:
\begin{equation}
x = - \frac{\hat a_1 (18 \hat a_1 + \hat a_2 + 18 \hat a_2 q)}{36 \hat a_2}\left(1 + 6 \frac{\hat a_2}{\hat a_1} y^2 \right) \ . 
\end{equation} 

Certain generalizations of the BBBW class of solutions have also appeared \cite{bah,bah-d6}. It would be interesting to reduce these to solutions of IIA supergravity and verify that they fit in our classification of section \ref{generalanalysis}.

\bibliography{at}

\providecommand{\href}[2]{#2}\begin{thebibliography}{10}

\bibitem{gauntlett-martelli-sparks-waldram-ads5-IIB}
J.~P. Gauntlett, D.~Martelli, J.~Sparks, and D.~Waldram, ``Supersymmetric
  {AdS$_5$} solutions of type {IIB} supergravity,'' {\em Class. Quant. Grav.}
  {\bf 23} (2006) 4693--4718,
\href{http://arXiv.org/abs/hep-th/0510125}{{\tt hep-th/0510125}}.

\bibitem{Pilch:2000ej}
K.~Pilch and N.~P. Warner, ``{A New supersymmetric compactification of chiral
  IIB supergravity},'' {\em Phys.Lett.} {\bf B487} (2000) 22--29,
\href{http://arXiv.org/abs/hep-th/0002192}{{\tt hep-th/0002192}}.

\bibitem{gauntlett-martelli-sparks-waldram-M}
J.~P. Gauntlett, D.~Martelli, J.~Sparks, and D.~Waldram, ``{Supersymmetric
  AdS$_5$ solutions of M theory},'' {\em Class.Quant.Grav.} {\bf 21} (2004)
  4335--4366,
\href{http://arXiv.org/abs/hep-th/0402153}{{\tt hep-th/0402153}}.

\bibitem{lin-lunin-maldacena}
H.~Lin, O.~Lunin, and J.~M. Maldacena, ``{Bubbling AdS space and 1/2 BPS
  geometries},'' {\em JHEP} {\bf 0410} (2004) 025,
\href{http://arXiv.org/abs/hep-th/0409174}{{\tt hep-th/0409174}}.

\bibitem{maldacena-nunez}
J.~M. Maldacena and C.~N{\'u}{\~n}ez, ``Supergravity description of field
  theories on curved manifolds and a no-go theorem,'' {\em Int. J. Mod. Phys.}
  {\bf A16} (2001) 822--855,
\href{http://arXiv.org/abs/hep-th/0007018}{{\tt hep-th/0007018}}.

\bibitem{gaiotto-maldacena}
D.~Gaiotto and J.~Maldacena, ``{The Gravity duals of ${\cal N}=2$
  superconformal field theories},'' {\em JHEP} {\bf 1210} (2012) 189,
\href{http://arXiv.org/abs/0904.4466}{{\tt 0904.4466}}.

\bibitem{bah-beem-bobev-wecht}
I.~Bah, C.~Beem, N.~Bobev, and B.~Wecht, ``{Four-Dimensional SCFTs from
  M5-Branes},'' {\em JHEP} {\bf 1206} (2012) 005,
\href{http://arXiv.org/abs/1203.0303}{{\tt 1203.0303}}.

\bibitem{afrt}
F.~Apruzzi, M.~Fazzi, D.~Rosa, and A.~Tomasiello, ``{All AdS$_7$ solutions of
  type II supergravity},'' {\em JHEP} {\bf 1404} (2014) 064,
\href{http://arXiv.org/abs/1309.2949}{{\tt 1309.2949}}.

\bibitem{blaback-danielsson-junghans-vanriet-wrase-zagermann}
J.~{Bl\aa b\"ack}, U.~H. Danielsson, D.~Junghans, T.~Van~Riet, T.~Wrase, and
  M.~Zagermann, ``{The problematic backreaction of SUSY-breaking branes},''
  {\em JHEP} {\bf 1108} (2011) 105,
\href{http://arXiv.org/abs/1105.4879}{{\tt 1105.4879}}.

\bibitem{gautason-junghans-zagermann}
F.~F. Gautason, D.~Junghans, and M.~Zagermann, ``{Cosmological Constant, Near
  Brane Behavior and Singularities},'' {\em JHEP} {\bf 1309} (2013) 123,
\href{http://arXiv.org/abs/1301.5647}{{\tt 1301.5647}}.

\bibitem{blaback-danielsson-junghans-vanriet-wrase-zagermann-2}
J.~{Bl\aa b\"ack}, U.~H. Danielsson, D.~Junghans, T.~Van~Riet, T.~Wrase, and
  M.~Zagermann, ``{(Anti-)Brane backreaction beyond perturbation theory},''
  {\em JHEP} {\bf 1202} (2012) 025,
\href{http://arXiv.org/abs/1111.2605}{{\tt 1111.2605}}.

\bibitem{junghans-schmidt-zagermann}
D.~Junghans, D.~Schmidt, and M.~Zagermann, ``{Curvature-induced Resolution of
  Anti-brane Singularities},''
\href{http://arXiv.org/abs/1402.6040}{{\tt 1402.6040}}.

\bibitem{gaiotto-t-6d}
D.~Gaiotto and A.~Tomasiello, ``{Holography for $(1,0)$ theories in six
  dimensions},'' {\em JHEP} {\bf 1412} (2014) 003,
\href{http://arXiv.org/abs/1404.0711}{{\tt 1404.0711}}.

\bibitem{hanany-zaffaroni-6d}
A.~Hanany and A.~Zaffaroni, ``{Branes and six-dimensional supersymmetric
  theories},'' {\em Nucl.Phys.} {\bf B529} (1998) 180--206,
\href{http://arXiv.org/abs/hep-th/9712145}{{\tt hep-th/9712145}}.

\bibitem{brunner-karch}
I.~Brunner and A.~Karch, ``{Branes at orbifolds versus Hanany--Witten in six
  dimensions},'' {\em JHEP} {\bf 9803} (1998) 003,
\href{http://arXiv.org/abs/hep-th/9712143}{{\tt hep-th/9712143}}.

\bibitem{intriligator-6d}
K.~A. Intriligator, ``{RG fixed points in six dimensions via branes at orbifold
  singularities},'' {\em Nucl.Phys.} {\bf B496} (1997) 177--190,
\href{http://arXiv.org/abs/hep-th/9702038}{{\tt hep-th/9702038}}.

\bibitem{intriligator-6d-II}
K.~A. Intriligator, ``{New string theories in six dimensions via branes at
  orbifold singularities},'' {\em Adv.Theor.Math.Phys.} {\bf 1} (1998)
  271--282,
\href{http://arXiv.org/abs/hep-th/9708117}{{\tt hep-th/9708117}}.

\bibitem{heckman-morrison-vafa}
J.~J. Heckman, D.~R. Morrison, and C.~Vafa, ``{On the Classification of 6D
  SCFTs and Generalized ADE Orbifolds},''
\href{http://arXiv.org/abs/1312.5746}{{\tt 1312.5746}}.

\bibitem{dhtv}
M.~Del~Zotto, J.~J. Heckman, A.~Tomasiello, and C.~Vafa, ``{6d Conformal
  Matter},'' {\em JHEP} {\bf 1502} (2015) 054,
\href{http://arXiv.org/abs/1407.6359}{{\tt 1407.6359}}.

\bibitem{rota-t}
A.~Rota and A.~Tomasiello, ``{AdS$_4$ compactifications of AdS$_7$ solutions in
  type II supergravity},''
\href{http://arXiv.org/abs/1502.06622}{{\tt 1502.06622}}.

\bibitem{myers}
R.~C. Myers, ``Dielectric-branes,'' {\em JHEP} {\bf 12} (1999) 022,
\href{http://arXiv.org/abs/hep-th/9910053}{{\tt hep-th/9910053}}.

\bibitem{polchinski-strassler}
J.~Polchinski and M.~J. Strassler, ``{The string dual of a confining
  four-dimensional gauge theory},''
\href{http://arXiv.org/abs/hep-th/0003136}{{\tt hep-th/0003136}}.

\bibitem{gmpt2}
M.~Gra{\~n}a, R.~Minasian, M.~Petrini, and A.~Tomasiello, ``Generalized
  structures of {${\cal N}=1$} vacua,'' {\em JHEP} {\bf 11} (2005) 020,
\href{http://arXiv.org/abs/hep-th/0505212}{{\tt hep-th/0505212}}.

\bibitem{gabella-gauntlett-palti-sparks-waldram}
M.~Gabella, J.~P. Gauntlett, E.~Palti, J.~Sparks, and D.~Waldram, ``{AdS$_5$
  Solutions of Type IIB Supergravity and Generalized Complex Geometry},'' {\em
  Commun.Math.Phys.} {\bf 299} (2010) 365--408,
\href{http://arXiv.org/abs/0906.4109}{{\tt 0906.4109}}.

\bibitem{anderson-beem-bobev-rastelli}
M.~T. Anderson, C.~Beem, N.~Bobev, and L.~Rastelli, ``{Holographic
  Uniformization},'' {\em Commun.Math.Phys.} {\bf 318} (2013) 429--471,
\href{http://arXiv.org/abs/1109.3724}{{\tt 1109.3724}}.

\bibitem{bah-beem-bobev-wecht0}
I.~Bah, C.~Beem, N.~Bobev, and B.~Wecht, ``{AdS/CFT Dual Pairs from M5-Branes
  on Riemann Surfaces},'' {\em Phys.Rev.} {\bf D85} (2012) 121901,
\href{http://arXiv.org/abs/1112.5487}{{\tt 1112.5487}}.

\bibitem{itsios-nunez-sfetsos-thompson}
G.~Itsios, C.~N{\'u}{\~n}ez, K.~Sfetsos, and D.~C. Thompson, ``{On Non-Abelian
  T-Duality and new ${\cal N}=1$ backgrounds},'' {\em Phys.Lett.} {\bf B721}
  (2013) 342--346,
\href{http://arXiv.org/abs/1212.4840}{{\tt 1212.4840}}.

\bibitem{witten-mqcdN2}
E.~Witten, ``{Solutions of four-dimensional field theories via M theory},''
  {\em Nucl.Phys.} {\bf B500} (1997) 3--42,
\href{http://arXiv.org/abs/hep-th/9703166}{{\tt hep-th/9703166}}.

\bibitem{aharony-berdichevsky-berkooz}
O.~Aharony, L.~Berdichevsky, and M.~Berkooz, ``{4d ${\cal N}=2$ superconformal
  linear quivers with type IIA duals},'' {\em JHEP} {\bf 1208} (2012) 131,
\href{http://arXiv.org/abs/1206.5916}{{\tt 1206.5916}}.

\bibitem{gaiotto}
D.~Gaiotto, ``{${\cal N}=2$ dualities},'' {\em JHEP} {\bf 1208} (2012) 034,
\href{http://arXiv.org/abs/0904.2715}{{\tt 0904.2715}}.

\bibitem{t-reform}
A.~Tomasiello, ``{Reformulating Supersymmetry with a Generalized {Dolbeault}
  Operator},'' {\em JHEP} {\bf 02} (2008) 010,
\href{http://arXiv.org/abs/arXiv:0704.2613 [hep-th]}{{\tt arXiv:0704.2613
  [hep-th]}}.

\bibitem{lust-tsimpis}
D.~{L\"ust} and D.~Tsimpis, ``Supersymmetric {AdS$_4$} compactifications of
  {IIA} supergravity,'' {\em JHEP} {\bf 02} (2005) 027,
\href{http://arXiv.org/abs/hep-th/0412250}{{\tt hep-th/0412250}}.

\bibitem{magoo}
O.~Aharony, S.~S. Gubser, J.~M. Maldacena, H.~Ooguri, and Y.~Oz, ``{Large $N$
  field theories, string theory and gravity},'' {\em Phys. Rept.} {\bf 323}
  (2000) 183--386,
\href{http://arXiv.org/abs/hep-th/9905111}{{\tt hep-th/9905111}}.

\bibitem{sfetsos-thompson}
K.~Sfetsos and D.~C. Thompson, ``{On non-abelian T-dual geometries with Ramond
  fluxes},'' {\em Nucl.Phys.} {\bf B846} (2011) 21--42,
\href{http://arXiv.org/abs/1012.1320}{{\tt 1012.1320}}.

\bibitem{sfetsos-thompson2}
K.~Sfetsos and D.~C. Thompson, ``{New ${\cal N} = 1$ supersymmetric ${\rm
  AdS}_5$ backgrounds in Type IIA supergravity},'' {\em JHEP} {\bf 1411} (2014)
  006,
\href{http://arXiv.org/abs/1408.6545}{{\tt 1408.6545}}.

\bibitem{reidedwards-stefanski}
R.~Reid-Edwards and j.~Stefanski, B., ``{On Type IIA geometries dual to ${\cal
  N}=2$ SCFTs},'' {\em Nucl.Phys.} {\bf B849} (2011) 549--572,
\href{http://arXiv.org/abs/1011.0216}{{\tt 1011.0216}}.

\bibitem{bah}
I.~Bah, ``{Quarter-BPS AdS$_{5}$ solutions in M-theory with a $T^{2}$ bundle
  over a Riemann surface},'' {\em JHEP} {\bf 1308} (2013) 137,
\href{http://arXiv.org/abs/1304.4954}{{\tt 1304.4954}}.

\bibitem{bah-d6}
I.~Bah, ``{AdS$_5$ solutions from M5-branes on Riemann surface and D6-branes
  sources},''
\href{http://arXiv.org/abs/1501.06072}{{\tt 1501.06072}}.

\end{thebibliography}
\bibliographystyle{at}

\end{document}